&pdflatex format=pdflatex
\documentclass[prd,preprint,eqsecnum,nofootinbib,amsmath,amssymb,preprintnumbers,tightenlines,longbibliography]{revtex4-1}

\usepackage[mathscr]{eucal}
\usepackage{mathrsfs}
\usepackage{hyperref}
\usepackage{graphicx}
\usepackage{bm}

\usepackage{graphicx}
\usepackage{bm}

\def\Lo{{L_o}}
\def\Pr{{\mathcal P}}

\def\RR{\mathfrak{R} }
\def\dd{{\rm d}}

\def\M{{\cal M}}
\def\x{{\bm x}}
\def\y{{\bm y}}

\def\z{{\bm z}}

\def\v{{\bm v}}

\def\p{{\bm p}}
\def\k{{\bm k}}
\def\q{{\bm q}}

\def\v{{\bm v}}

\def\twothirds{{\textstyle\frac{2}{3}}}
\def\third{{\textstyle\frac{1}{3}}}

\def\x{{\bm x}}

\def\kn{{k} }

\def\st{\begin{equation}}
\def\stp{\end{equation}}
\def\bg{\begin{eqnarray}}
\def\nd{\end{eqnarray}}
\def\Eq#1{Eq.~(\ref{#1})}
\def\app#1{Appendix~\ref{#1}}
\def\Fig#1{Fig.~\ref{#1}}
\def\Sect#1{Section~\ref{#1}}
\def\N{{\mathcal N}}

\def\llangle{\left\langle}
\def\rrangle{\right\rangle}

\def\Eq#1{Eq.~(\ref{#1})}
\def\app#1{Appendix~\ref{#1}}
\def\Fig#1{Fig.~\ref{#1}}
\def\Sect#1{Section~\ref{#1}}
\def\Ref#1{Ref.~\cite{#1}}

\def\gsim{\mbox{~{\protect\raisebox{0.4ex}{$>$}}\hspace{-1.1em}
	{\protect\raisebox{-0.6ex}{$\sim$}}~}}
\def\lsim{\mbox{~{\protect\raisebox{0.4ex}{$<$}}\hspace{-1.1em}
	{\protect\raisebox{-0.6ex}{$\sim$}}~}}

\def\nott#1{\setbox0=\hbox{$#1$}                
   \dimen0=\wd0                                
   \setbox1=\hbox{/} \dimen1=\wd1               
   \ifdim\dimen0>\dimen1                        
      \rlap{\hbox to \dimen0{\hfil/\hfil}}      
      #1                                        
   \else                                        
      \rlap{\hbox to \dimen1{\hfil$#1$\hfil}}   
      /                                         
   \fi}                                         %

\advance\parskip 1.9pt
\advance\voffset -0.2in

\begin{document}
\title{The Wake of a Heavy Quark in Non-Abelian Plasmas :  Comparing Kinetic Theory and the  AdS/CFT Correspondence}
\author{Juhee Hong}
\affiliation{Department of Physics and Astronomy, Stony Brook University, 
Stony Brook, NY 11794-3800, United States}
\author{Derek Teaney}
\affiliation{Department of Physics and Astronomy, Stony Brook University, 
Stony Brook, NY 11794-3800, United States}
\author{Paul M. Chesler}
\affiliation{Center for Theoretical Physics, MIT, 
Cambridge, MA 02139, United States}
\date{\today}

\begin{abstract}
%
We compute the non-equilibrium stress tensor induced by a heavy quark 
moving through weakly coupled QCD plasma at the speed of light and  compare the  result
to $\N=4$ Super Yang Mills theory at strong coupling.
The  QCD Boltzmann equation is reformulated as a Fokker-Planck equation in a leading log approximation which is used to compute the induced stress.
The transition from non-equilibrium at short distances  to equilibrium at large distances is analyzed with first and second order hydrodynamics.

Even after accounting for the obvious differences in shear lengths,
the strongly coupled theory  is significantly better described
by hydrodynamics at sub-asymptotic distances. 
We argue that this difference between the kinetic and AdS/CFT theories is
related to the second order hydrodynamic coefficient $\tau_\pi$. $\tau_\pi$ is
numerically large in units of the  shear length for theories based on the
Boltzmann equation. 

\end{abstract}

\maketitle

\section{Introduction}

The goal of the relativistic heavy ion programs at  RHIC and the LHC 
is to create and study the properties of the Quark Gluon Plasma (QGP). 
Since the real time dynamics  can not be directly probed with 
the Lattice QCD, the transport properties of QGP are of particular interest.
There is a consensus  that the
experimental results on collective flow imply that the shear viscosity to
entropy ratio of the Quark Gluon Plasma is remarkably small 
\cite{*[{For a
review of the degree of consensus and references see:}] Teaney:2009qa}
\st
 \frac{\eta}{s} \sim \frac{1 \leftrightarrow 5}{4\pi}  \, .
\stp
Although it is difficult to reconcile this experimental ratio with a quasi-particle
picture  of the plasma,   the success of resummed perturbation
theory at describing lattice data on the pressure at somewhat higher temperatures 
 suggests
that a quasi-particle picture might provide a qualitative guide to the plasma
dynamics and form a basis for further approximation schemes  
\cite{Blaizot:2003iq,Andersen:2011sf,Andersen:2011ug}.

However, 
in  $\N=4$ Super Yang Mills (SYM) theory at strong coupling (and large $N_c$)
gauge-gravity duality can be used to compute the  
shear to entropy ratio  exactly \cite{Policastro:2001yc,Kovtun:2004de},
\st
 \frac{\eta}{s}= \frac{1}{4\pi} \, .
\stp
The tantalizing similarity between the experimental ratio 
and this celebrated theoretical result hints that a strong 
coupling limit (without quasi-particles)  
might provide a better starting point for understanding
the plasma dynamics. 
At the very least, the strongly coupled $\N=4$ theory is an analytically tractable 
limit that provides a useful foil to perturbative calculations based
on a quasi-particle description.

The goal of this work is to compare the steady state response of non-abelian plasma at weak and strong coupling to an
infinitely heavy quark probe  moving at the speed of light.
This is the simplest setup where the plasma response to an energetic probe 
can be analyzed in detail
\cite{Stoecker:2004qu,CasalderreySolana:2004qm,Gubser:2007ga,Yarom:2007ap,Chesler:2007an,Chesler:2007sv,Neufeld:2008hs}.  
At long distances, the non-equilibrium disturbance produced by the 
heavy quark probe thermalizes and forms a Mach cone and diffusion wake.
The
original motivation for investigating the Mach cone 
was the observation of an unusual structure in
measured two particle correlations \cite{Adare:2008cqb,Abelev:2008nda}.  Today, after the analysis of Alver and Roland \cite{Alver:2010gr} and others 
\cite{Takahashi:2009na,Sorensen:2010zq,Lin:2004en,Ma:2006fm},  these unusual correlations are
understood as the hydrodynamic response to fluctuations in the initial geometry
and not as the medium  response to an  energetic probe.
(The Mach cone picture also dramatically fails to explain current
measurements in several ways -- see for example \Ref{Chaudhuri:2005vc} and  the conclusions of
\Ref{CasalderreySolana:2006sq}.) 
  The goal 
of this manuscript is not to explain current measurements, but rather
to examine the differences between weak and strong coupling, and to 
study the  approach to hydrodynamics in both cases.
Although the current paper has no immediate phenomenological goals, 
 the medium response to energetic partons is 
currently being studied by all the experimental 
collaborations in various ways \cite{[{See for example the recent analysis by the CMS collaboration: }] Chatrchyan:2011sx}.  
Thus, this   calculation, which 
analyzes the ``jet" medium interaction precisely and  
determines a source  for hydrodynamics through second order  in 
the gradient expansion,  may  be useful for phenomenology in further studies.
 
In the strongly coupled theory
the  stress tensor
induced by a finite velocity heavy quark was computed  using the
AdS/CFT correspondence \cite{Chesler:2007an, Gubser:2007ga}.  The approach to
hydrodynamics was analyzed  as well as the short distance behavior
\cite{Chesler:2007sv,Yarom:2007ap,Gubser:2007nd}.  In particular  we will largely follow (and to a certain extent
extend) the hydrodynamic analysis of \Ref{Chesler:2007sv}  to determine a
hydrodynamic source through second order in the gradient expansion for the
kinetic and strongly coupled theories.  
In the AdS/CFT calculation the lightlike $v\rightarrow c$ limit was
not analyzed due to various technical complications. (Here and below $v$
is the velocity of the heavy quark.) As discussed in \Sect{adscft}, it is
possible to set $v =c$ throughout the calculation by choosing a different set
of gauge invariants. 

At weak coupling the hydrodynamic source has not been computed. Nevertheless
the appropriate source for kinetic theory was determined in
\Ref{Neufeld:2008hs},  and several estimates have been given for how this
kinetic source
is transformed through the relaxation process to hydrodynamics  \cite{Neufeld:2008dx}.
We have simplified the source for kinetic theory considerably and 
determined  the plasma response  at large distances by solving the linearized 
kinetic theory. After 
comparing the hydrodynamic solution at large distances to 
the full (leading-log) kinetic theory results, the 
appropriate source at each order in the hydrodynamic expansion can be computed.
This part of the calculation employs a computer code developed 
by us to determine  spectral functions at finite $\omega$ and $k$ \cite{Hong:2010at}. 
As a by-product of these spectral functions we determined  the hydrodynamic
transport coefficients that appear through second order in 
the gradient expansion in a leading log approximation.  These parameters
will be needed below to precisely determine the hydrodynamic source through
second order.

This work is limited to the analysis of the kinetics  for a single heavy
quark moving from past infinity.  It would be quite interesting  to
follow
the evolution of a  parton shower initiated at time $t=0$ 
and the subsequent hydrodynamic response at late times.  
Although this transition has not been
worked out,  several of the most important ingredients have already been
clarified \cite{Neufeld:2009ep,Neufeld:2011yh,Arnold:2009ik}.   We hope to
address the thermalization of  full parton showers in future work.

\section{Preliminaries}

We  consider an infinitely heavy quark with $v\simeq 1$ moving  through a stationary high temperature plasma from past infinity.
We will calculate the medium response in two model theories -- pure glue QCD  at asymptotically weak coupling  and $\N=4$ SYM at asymptotically strong coupling. Both theories are conformal in this limit and therefore
the background stress
tensor  takes  the characteristic form
\st
  T^{\mu\nu}_o = {\rm diag}(e,\Pr, \Pr, \Pr) \,  , 
\qquad \mbox{with} \qquad e = 3\Pr.
\stp 
The heavy quark moves in the $\hat{\z}$ direction 
and imparts energy and 
momentum to the plasma, which ultimately  induces
a non-equilibrium response,  $\delta T^{\mu\nu}$.
The non-equilibrium stress tensor
$\delta T^{00}$ and $\delta T^{0z}$ are functions of cylindrical 
and comoving coordinates $x_T$ and
$x_L$ where  
\st
\label{cylindrical1}
x_{T} = \sqrt{x^2 + y^2} \, , \qquad \varphi_{r} = {\rm atan2}(y,x) \, , \qquad \mbox{and} \qquad  x_L = z - v t \, . 
\stp
Rotational invariance around the $z$ axis determines  $(T^{0x},
T^{0y})$ in terms of $T^{0x_T}$  
\begin{align}
\label{cylindrical2}
   T^{0x}(t,\x) =& T^{0x_T}(x_L,x_T) \cos \varphi_r\, ,  \\
   T^{0y}(t,\x) =& T^{0x_T}(x_L,x_T) \sin \varphi_r\, .  
\end{align}

\subsection{Kinetic Theory with a Heavy Quark Probe}

At weak coupling kinetic theory determines the response of the plasma to the heavy quark probe.
To determine this response we linearize the Boltzmann equation for $f(t,\x,\p)=n_p+\delta f(t,\x,\p)$ 
with $n_p=1/(e^{p/T_o}-1)$ and will restrict the calculation to  pure glue QCD in a leading log-approximation for simplicity\footnote{ 
Including  quarks would  only lead  to minor
changes  to our results as can be seen from Fig.~4 of \Ref{Hong:2010at}. }.
The Boltzmann equation in this limit reads 
\bg
\label{be}
\left(\partial_t+\v_p\cdot\partial_{\x}\right)\delta f(t,\x,\p)
=C[f,\p] +  S(t,\x, \p) \, ,
\nd
where $\v_\p =\hat{\p}$ and $S(t,\x,\p)$ is the (to be discussed)  source 
of non-equilibrium gluons produced by the heavy quark moving through the plasma.
In a leading $\log(T/m_D)$ approximation, the linearized collision integral   simplifies to a momentum diffusion equation  supplemented by gain terms \cite{Hong:2010at}
\bg
C[f,\p]&=&T\mu_A \frac{\partial}{\partial \p^i}\left(  n_p(1 + n_p) 
\frac{\partial} {\partial \p^i} 
\left[ \frac{\delta f }{n_p (1 + n_p)} \right] \right) 
+ \mbox{gain terms} \, , \nonumber\\
\mu_A &\equiv & \frac{g^2C_A m_D^2 }{8 \pi} \log\left(\frac{T}{m_D}\right) \, .
\nd  
Here the Debye mass for a pure glue theory is 
\st
    m_{D}^2 = 2 g^2 T_{A} \int \frac{\dd^3\p}{(2\pi)^3 }  \frac{n_p (1+ n_p)}{T} = \frac{g^2 T^2  }{3} N_c \, ,
\stp
where $T_{A}=N_c$ is the trace normalization of the adjoint representation.
The diffusion equation should be solved with absorptive boundary conditions
at $p=0$ so the number of particles is not conserved during the evolution \cite{Arnold:2006fz}. 
Thus the microscopic theory encoded by this diffusion equation is conformal, and the only conserved quantities are   energy and momentum.  

The gain terms are responsible for energy and momentum conservation.
Specifically, the energy and momentum  
that is transferred (per time, per degree of freedom, per volume) 
to the non-equilibrium excess $\delta f$
 by equilibrium bath is
\begin{align}
 \frac{\dd E}{\dd t} \equiv&   - T\mu_A \int  \frac{\dd^3\p}{(2\pi)^3 } \,  
   n_p(1 + n_p)  \, \hat \p   \cdot \frac{\partial }{\partial \p } \left[ \frac{\delta f } {n_p (1 + n_p) } \right] \, , \\
 \frac{\dd{\bm P}}{\dd t}  \equiv& - T\mu_A \int \frac{\dd^3\p}{(2\pi)^3 }\,  
   n_p(1 + n_p)  \, \frac{\partial }{\partial \p } \left[ \frac{\delta f } {n_p (1 + n_p) } \right] \, ,
\end{align}
as can easily be found by integrating  both sides of \Eq{be} without the
source.
This energy and momentum transfer by the bath
drives additional particles away from equilibrium and ultimately
fixes the structure of the gain terms:
\begin{align}
\label{gain_terms}
\mbox{gain terms} =  
  \frac{1}{\xi_B} 
\left[ \frac{1}{p^2}\frac{\partial}{\partial p}p^2 n_p(1 + n_p)\right] \frac{\dd E}{\dd t} 
  + \frac{1}{\xi_B} \left[\frac{\partial}{\partial \p} n_p (1 + n_p)  \right]  \cdot \frac{ \dd{\bm P}}{\dd t}  \; ,
\end{align}
where for subsequent use we have defined
\begin{align}
\label{xis}
\xi_B \equiv \int \frac{\dd^3 \p}{(2\pi)^3 } n_p(1 + n_p) = \frac{T^3}{6}  \, .
\end{align}
With the gain terms it is easy to verify that energy and momentum are
conserved.  Previously we analyzed the linear response of
this system of equations 
and determined the hydrodynamic plasma parameters in terms of $\mu_{A}$ \cite{Hong:2010at}
\begin{align}
\frac{\eta }{e + P} =& 0.4613\, \frac{T}{\mu_A} \, ,  \\
\frac{\tau_\pi}{\eta/sT} =& 6.32 \, .
\end{align}
Since the microscopic dynamics is conformal, linearized, and only 
conserves energy and  momentum (and not particle number),  $\tau_\pi$ is the only second order
hydrodynamic coefficient that appears to this order.  If there are additional
conserved quantities and the dynamics is not conformal  then there 
are a multitude of coefficients that appear 
at second order -- see for example \cite{Romatschke:2009kr,Betz:2010cx}. Further 
it must be emphasized that these second order transport coefficients
are insufficient to describe the decay of initial transients  (non-hydrodynamic modes) \cite{Betz:2010cx}.

The shear viscosity naturally agrees with prior results \cite{Baym:1990uj,Arnold:2000dr}.
The fact that $\tau_\pi$ is somewhat large $\sim 6$ compared to
the viscous length is a generic result of kinetic theory \cite{Baier:2007ix,York:2008rr}. 
Finally, we note that $\mu_{A}$  records the transverse momentum broadening of a bath particle due to 
the soft scatterings,  and is related to the soft
part of jet-quenching $\hat{q}$
parameter in a leading $T/m_D$ approximation \cite{Arnold:2008vd}, $\hat{q}_{\rm soft}/2 = 2
T\mu_{A}$. 
Thus, the leading-log limit provides a concrete 
relation between $\eta/s$ and $\hat q$.

We now  analyze how a heavy quark disturbs this system in  the same
leading log approximation scheme.
 The leading log energy loss of the   
heavy quark was computed long ago by Braaten and Thoma \cite{Braaten:1991jj,Braaten:1991we}, 
with the result that the momentum transferred to the 
medium  per time (\emph{i.e.} minus the drag-force) is
\st
\label{drag}
  \frac{\dd p^\mu}{\dd t}  = \left(\frac{\dd E}{\dd t} , \frac{\dd \p}{\dd t} \right) = \mu_{F}(v) \, \Big(v^2, \v \Big) \, , 
\stp
where  $\mu_{F}(v)$ is the drag coefficient in a leading
log approximation  
\begin{align}
\label{mudef_kt}
\mu_F(v) =& 
 \frac{g^2 C_F m_D^2}{8\pi} \log\left(\frac{T}{m_D} \right) 
\left(\frac{1}{v^2} -\frac{1-v^2}{2v^3} \log\left(\frac{1 + v}{1-v}\right) \right)   \, ,   \\
          \Rightarrow & 
 \frac{g^2 C_F m_D^2}{8\pi} \log\left(\frac{T}{m_D} \right)  \, .
\end{align} 
In the last line we have taken the $v \rightarrow 1$ limit of relevance to this
work.   
As discussed more completely below, we have implicitly taken the coupling to zero  before 
taking this limit so that radiative
energy loss can be neglected.

The drag force arises as  equilibrium gluons from the bath scatter off  
the heavy quark probe and are driven out of equilibrium 
by the scattering  process. 
This scattering produces 
a source of non-equilibrium gluons located at the 
position of the quark, 
\[
  S(t, \x, \p) =  S(\p) \, \delta^3(\x - \v t)  \, .
\]
\app{source_app} analyzes this scattering process ($g+ Q \rightarrow g + Q$) and determines
the appropriate momentum space source, $S(\p)$. 
In the limit $v=1$ the source has a simple form involving only two spherical harmonics 
\bg
\label{source}
S(\p)&=&
\mu_F \, \frac{n_p(1+n_p)}{2d_A \xi_B} \left[-\frac{2}{p}+\frac{(1+2n_p)}{T}
+ \frac{(1+2n_p)}{T}  \, \hat{\p} \cdot {\hat \v} \right],  \qquad \mbox{for $v=1$ \, . } 
\nd
With this source it is straightforward to 
integrate 
$2d_A\int \frac {\dd^3\p}{(2\pi)^3} p^{\mu}$ over \Eq{be} and 
to verify that the stress tensor
satisfies
\st
\label{micro-eom}
    \partial_{\mu} \, \delta T^{\mu\nu}  = \frac{ \dd p^{\nu} }{\dd t} \,  
\delta^{3}(\x- \v t)  \, .
\stp

Our strategy to determine the energy-momentum tensor  is the following. 
We take the Fourier transform with respect to $\x$ of the kinetic 
equation in \Eq{be} 
\st
(-i\omega+i\v_p\cdot\k)\delta f(\omega,\k,\p)
=C[\delta f,\p]+ 2\pi S(\p)\delta(\omega-\v\cdot\k) \, ,
\stp
and  solve the Boltzmann equation in Fourier  space. 
The technology to do this is based on simple matrix inversion as was documented in \Ref{Hong:2010at}.
Then we calculate the stress energy tensor in  Fourier space using kinetic theory. 
By Fourier transforming the stress tensor back to
coordinate space, we can determine the energy and momentum density distributions. 
Additional details  about this procedure are 
given in \app{details}.

\subsection{AdS/CFT with a heavy quark probe }
\label{adscft}

To describe the response of the $\N=4$ plasma to a heavy quark probe
we will follow the notations and conventions of Refs.~\cite{Chesler:2007an,Athanasiou:2010pv} which should be referred to for all details.
Briefly, a heavy quark is described in ${\rm AdS}_{5}$
with a trailing string. 
 The 
energy and momentum 
gained by the medium as the heavy quark traverses the
plasma is again parameterized with the drag coefficient $\mu_F(v)$
\st
  \frac{\dd p^\mu}{\dd t}  = \left(\frac{\dd E}{\dd t} , \frac{\dd \p}{\dd t} \right) = \mu_{F}(v)  \Big( v^2, \v \Big) \, . 
\stp
This coefficient is found by determining  the energy and momentum flowing down the string into  the black hole
 \cite{Herzog:2006gh,[{}] [{. This work was limited to the non-relativistic limit.}] CasalderreySolana:2006rq,Gubser:2006bz} 
\st
\label{mudef_ads}
   \mu_{F}(v) = \frac{\pi}{2}  \frac{\sqrt{\lambda} T^2 }{\sqrt{1-v^2} } \, .
\stp
As in the weakly coupled case,
the stress tensor in the strongly coupled $\N=4$ theory satisfies
\Eq{micro-eom} with the energy-momentum transfer rates given by the 
corresponding strongly coupled formulas.
The deposited energy and  momentum leads ultimately 
to a hydrodynamic response in the strongly
coupled theory.  The linearized hydrodynamic parameters through second order
are  \cite{Policastro:2001yc,Baier:2007ix,Bhattacharyya:2008jc} 
\begin{align}
  \frac{\eta}{e + P} =&  \frac{1}{4\pi T}  \, , \\
   \frac{\tau_\pi}{\eta/(e+P) }  =&   4  - 2 \log(2) \simeq 2.61 \, .
\end{align}


%
%

%



According to AdS/CFT duality \cite{Maldacena:1997re}, strongly coupled SYM plasma is dual to the $5d$ AdS-Schwarzschild
geometry, which has the metric
\begin{equation}
\label{metric}
ds^2 = \frac{L^2}{u^2} \left [ -f(u) dt^2 +  d\x^2 + \frac{du^2}{f(u)} \right] \, .
\end{equation}
Here $u$ is the radial coordinate of the AdS geometry with $u = 0$ corresponding to the boundary,
$L$ is the AdS curvature radius,   $f(u) = 1 - u^4/u_h^4$ with $u_h = 1/\pi T$, and  $T$  is the (Hawking) temperature of the plasma and dual geometry.

The addition of an infinitely massive fundamental quark to the SYM plasma is dual to 
the addition of a string to the AdS Schwarzschild geometry, with the string ending at $u = 0$ \cite{Karch:2003nh}.
The presence of the string perturbs the $5d$ geometry according to Einstein's equations and the near
boundary behavior of the metric perturbation encodes the changes in the SYM stress tensor due to the presence of the quark \cite{deHaro:2000xn}.

In the large $N$ limit the $5d$ gravitational constant $\kappa_5^2 \sim 1/N^2$  is small. Consequently the back-reaction 
of the string on the geometry can be treated perturbatively by solving the string equations in the background metric and 
subsequently computing the metric perturbations sourced by the string. 
Solving the string equations of motion leads to the well known trailing string profile \cite{Herzog:2006gh, Gubser:2006bz}
\begin{equation}
\bm x_{\rm string}(t,u) = \bm v \left [ t + \frac{u_h}{2} \left ( \tan^{-1} \frac{u}{u_h} + \frac{1}{2} \log \frac{u_h - u}{u_h + u} \right ) \right ].
\end{equation}
This string profile describes a quark moving at constant velocity $\bm v$
and has the $5d$ stress tensor
\begin{subequations}
\label{stress}
\begin{align}
t_{0i}   &= - v_i F, &
t_{ij}     &=  v_i v_j F, &
t_{00 } &= \frac{u^4 v^2 + uh^4 f }{uh^4} F, \\
t_{05}  &= - \frac{u^2 v^2}{u_h^2 f} F, &
t_{i 5}  &= \frac{u^2 v_i}{u_h^2 f} F, &
t_{55}  &= \frac{v^2 - f }{f^2} F,
\end{align}
\end{subequations}
where \begin{equation}
F = \frac{u \sqrt{\lambda}}{2 \pi L^3 \sqrt{1 - v^2}} \delta^3(\bm x - \bm x_{\rm string} ),
\end{equation}
and $\lambda$ is the 't Hooft coupling.

The five dimensional stress tensor of the string perturbs the background geometry.
The information contained in the linear metric perturbation sourced by the trailing string
can be conveniently packaged into fields which are invariant under infinitesimal diffeomorphisms \cite{Athanasiou:2010pv,Chesler:2007sv}.  Defining
\st
G_{MN} \equiv G_{MN}^{(0)} + \frac{L^2}{u^2} H_{MN} \, , 
\stp
where $G_{MN}^{(0)}$ is the background metric (\ref{metric})
and $H_{MN}$ is the perturbation, and introducing a space-time Fourier transform
\begin{equation}
H_{MN}(t,\bm x,u) = \int \frac{\dd \omega}{2 \pi} \frac{\dd^3 \q}{(2 \pi) ^3} H_{MN}(\omega,\bm q,u)e^{-i \omega t + i \bm q\cdot \bm x},
\end{equation}
we find two convenient diffeomorphism invariant fields \cite{Athanasiou:2010pv,Chesler:2007sv}

\begin{align}
Z_0 \equiv&  \frac{4 f}{\omega} q^i  H'_{0i} - \frac{4 f'}{\omega}q^i  H_{0i}  - \frac{2 f'  }{q^2} q^i q^jH_{ij} + 4 i f  q^i H_{i 5}
\nonumber   \\
& \qquad  \qquad \qquad \qquad 
- \frac{ (2 u q^2 {-} f')}{q^2} \left (q^2 \delta^{ij} {-}  q^i q^j  \right ) H_{ij} 
  + \frac{4 q^2 f}{i \omega} H_{05} - \frac{8 \kappa_5^2 f}{i \omega} t_{05}\,,
\\
    \label{vector}
    \bm  Z_1 \equiv &
    \left ( H'_{0i} - i \omega \,H_{i5} \right) \hat \epsilon^i_a \hat {\bm\epsilon}_{a} \, .
    \end{align}
Here sums over repeated indices are implied with $i, j$ running from 1 to 3 and $a$ running from 1 to 2, $'$ denotes differentiation with respect to $u$ and
\begin{equation}
\label{eps1}
    \hat {\bm\epsilon}_1 =
    \frac{q}{q_{\perp}} \,
    \hat {\bm q} \times (\hat {\bm v} \times \hat {\bm q}) \,,
\quad
    \hat {\bm\epsilon}_2 =
    \frac{q}{q_{\perp}} \, \hat {\bm v} \times \hat {\bm q} \,.
\end{equation}
The field $Z_0$ transforms as a scalar under rotations and the field $\bm Z_1$ transforms as a vector under rotations.%
\footnote
  {
  A complete set of gauge invariants also includes a field which transforms as a traceless symmetric tensor under
  rotations \cite{Chesler:2007sv}.  This tensor mode determines the spatial components of the SYM stress and is not necessary for 
  our purposes.
  }

The equations of motion for $Z_0$ and $\bm Z_1$ are straightforward but tedious to derive from the linearized Einstein
equations.  They read
\begin{equation}
\label{z0eqn}
Z_0'' + A_0 Z_0' + B_0 Z_0 = \kappa_5^2 S_0\,,
\end{equation}
where
\begin{align}
A_0 &\equiv -\frac{24 + 4 q^2 u^2 + 6 f + q^2 u^2 f - 30 f^2}{u f \left (u^2 q^2 + 6 - 6 f \right ) }\,,
\\
B_0 &\equiv \frac{\omega^2}{f^2} + \frac{  q^2 u^2(14 {-} 5 f {-} q^2 u^2) +18  (4{-}f{-} 3 f^2 )}{u^2 f \left ( q^2 u^2 +6 - 6f \right ) }\,,
\ \ \ \ \ \ \ \ 
\\ \label{generalsource}
S_0 &\equiv \frac{8 }{f} t'_{00}  
+ \frac{4 \left ( q^2 u^2 {+}6 {-} 6f \right ) }{3 u q^2 f} (q^2 \delta^{ij} {-} 3 q^i q^j ) t_{ij}  
\\ \nonumber
& \qquad 
+ \frac{8 i \omega}{f} t_{05}
+ \frac{8u \left[   q^2  \left(q^2 u^2{+}6\right) - f \left ( 12 q^2  {-} 9 f'' \right )\right]  }{3  f^2 \left(q^2 u^2-6 f+6\right)} t_{00} 
   -\frac{8 q^2 u}{3} t_{55}
- 8 i q^i t_{i5}\,.
\end{align}
and
\begin{equation}
\label{eqm1}
\bm Z''_1+A_1 \, \bm Z'_1 + B_1 \, \bm Z_1 = \kappa_5^2 \bm  S_1 \,,
\end{equation}
where
\begin{align}
A_1 &\equiv \frac{u f'-3 f}{ uf} \,,
\\ 
B_1 &\equiv \frac{3 f^2-u \left(u q^2+3 f'\right) f+u^2 \omega ^2}{u^2 f^2} \,,
\\
\label{S1general}
    \bm S_1 &\equiv   \frac{2}{f} \, \Big [ t'_{0i}+i \omega \, t_{i5} \Big ]   \hat \epsilon^i_a \hat {\bm\epsilon}_{a}.
\end{align}

We note that since the string stress tensor (\ref{stress}) only depends on time through the combination $\bm x - \bm v t$,
when Fourier transformed, the string stress tensor is proportional to $2 \pi \delta(\omega - \bm v \cdot \bm q)$.
Consequently the fields $Z_s$ are also proportional to $2 \pi \delta(\omega - \bm v \cdot \bm q)$.
Moreover, because the string stress tensor (\ref{stress}) is proportional to $1/\sqrt{1-v^2}$ and Eqs.~(\ref{z0eqn}) and (\ref{eqm1})
are linear, we define $Z_s = \tilde Z_s/\sqrt{1-v^2}$ and solve for $\tilde Z_s$ in the $v \to 1$ limit.

Under the assumption that the boundary geometry is flat, near the boundary the fields $Z_s$ have the 
asymptotic expansions 
\begin{equation}
\label{series}
\tilde Z_s(u) = \tilde Z_s^{(2)} u^2 +  \tilde  Z_s^{(3)} u^3 + \tilde  Z_s^{(4)} u^4  + \dots.
\end{equation}
The cubic expansion coefficients $\tilde  Z_s^{(3)}$ determine the SYM energy density $\delta T^{00}$ and  
the SYM energy flux $\delta T^{0i}$ via \cite{Athanasiou:2010pv,Chesler:2007sv}

\begin{align}
\label{eq:EfromZ3}
\delta T^{00}  =& -\frac{ L^3}{8 \kappa_5^2} \frac{1}{\sqrt{1-v^2}} \tilde Z^{(3)}_0 \, , \\ 
\label{eq:SfromZ}
\delta T^{0i} =&
   - \frac{L^3}{2 \kappa_5^2} \frac{1}{\sqrt{1-v^2}} \left [ ({\tilde Z_1^{(3)}})^i + \frac{\omega q^i}{4 q^2} (\tilde Z^{(3)}_0)^i \right ] - \frac{ i q^i \mu_F(v) v^2}{q^2}.
    \end{align}

For a given momentum $\bm q$, to determine the SYM energy density and energy flux we solve Eqs.~(\ref{z0eqn}) and (\ref{eqm1}) using pseudospectral 
methods.  At the boundary at $u = 0$ we impose the boundary condition that the fields have asymptotics of the form given 
in the series expansions (\ref{series}), which is tantamount to demanding that the boundary geometry is flat.  At the 
horizon at $u = u_h$ we impose the boundary condition of infalling waves.  This is tantamount to demanding that 
$\tilde Z_s \sim (u - u_h)^{-i \omega u_h/4}$ near the horizon.  With the $\tilde Z_s$ known, we can then extract the expansion coefficients $\tilde Z_s^{(3)}$ and 
construct the SYM energy density and energy flux from Eq.~(\ref{eq:EfromZ3}) and \Eq{eq:SfromZ} .

\section{Comparing AdS/CFT and Kinetic Theory} 
\label{comparison}

\begin{figure}
\begin{center}
\includegraphics[width=0.7\textwidth]{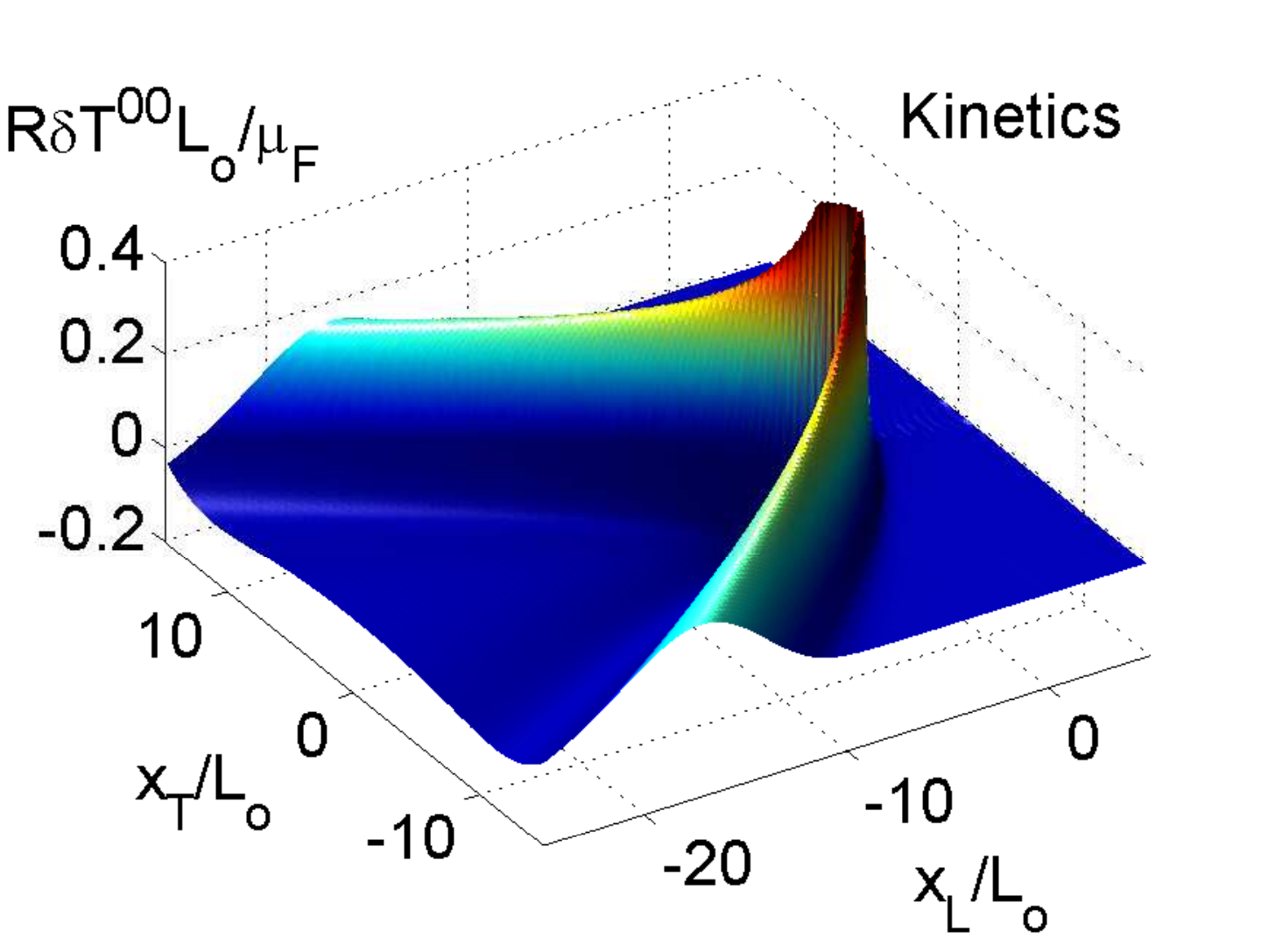}
\includegraphics[width=0.7\textwidth]{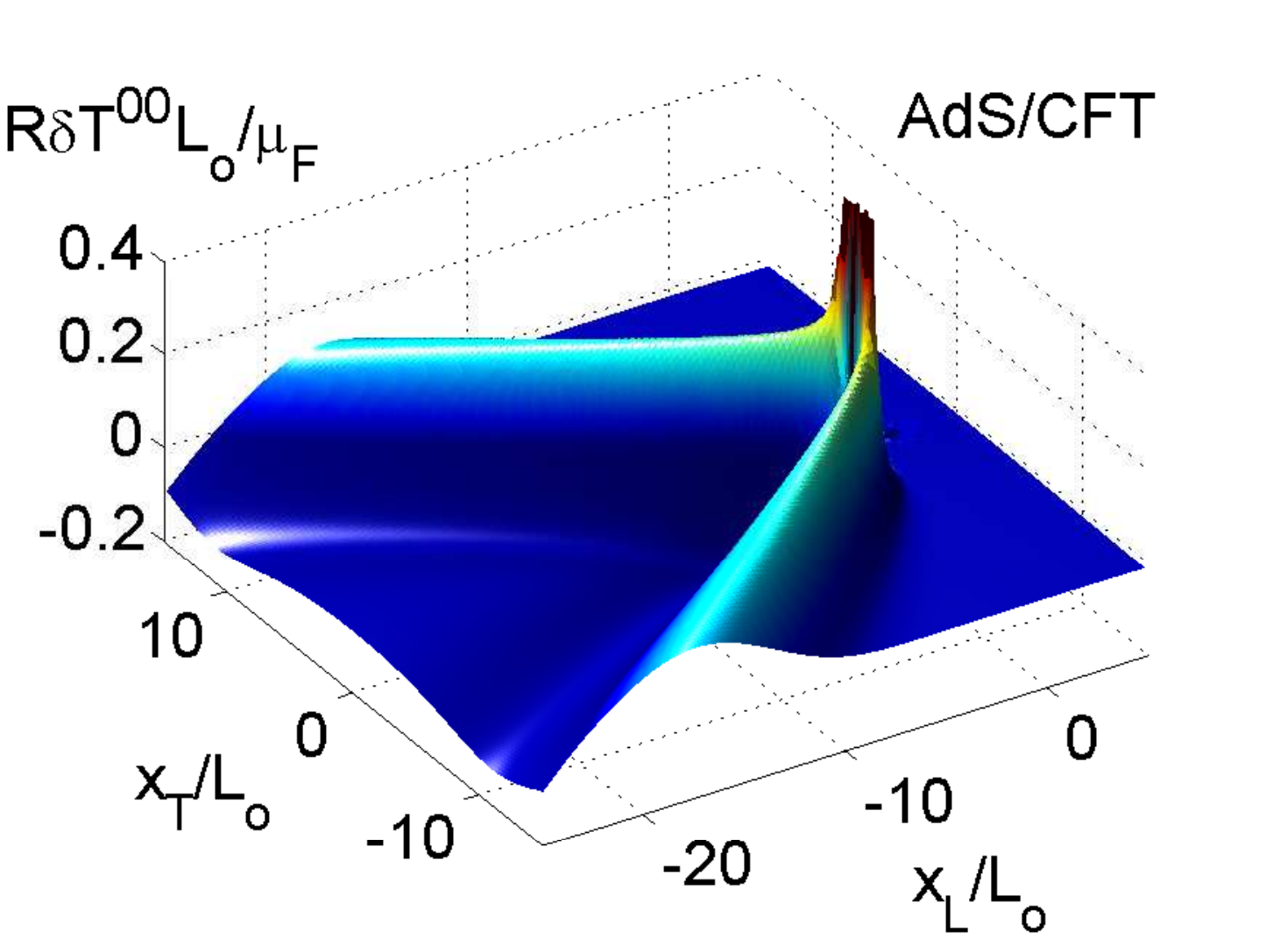}
\end{center}
\caption{
The energy density (in scaled units) times $R= \sqrt{x_T^2 + x_L^2}$ that is induced
by a heavy quark probe in (a) weakly coupled QCD and (b) strongly coupled
$\N=4$ SYM. 
Here $\Lo$ is the shear length and the $\mu_F(v)$ is the 
drag coefficient for each case (see text).
\label{EPfull}
}
\end{figure}

\begin{figure}
\begin{center}
\includegraphics[width=0.7\textwidth]{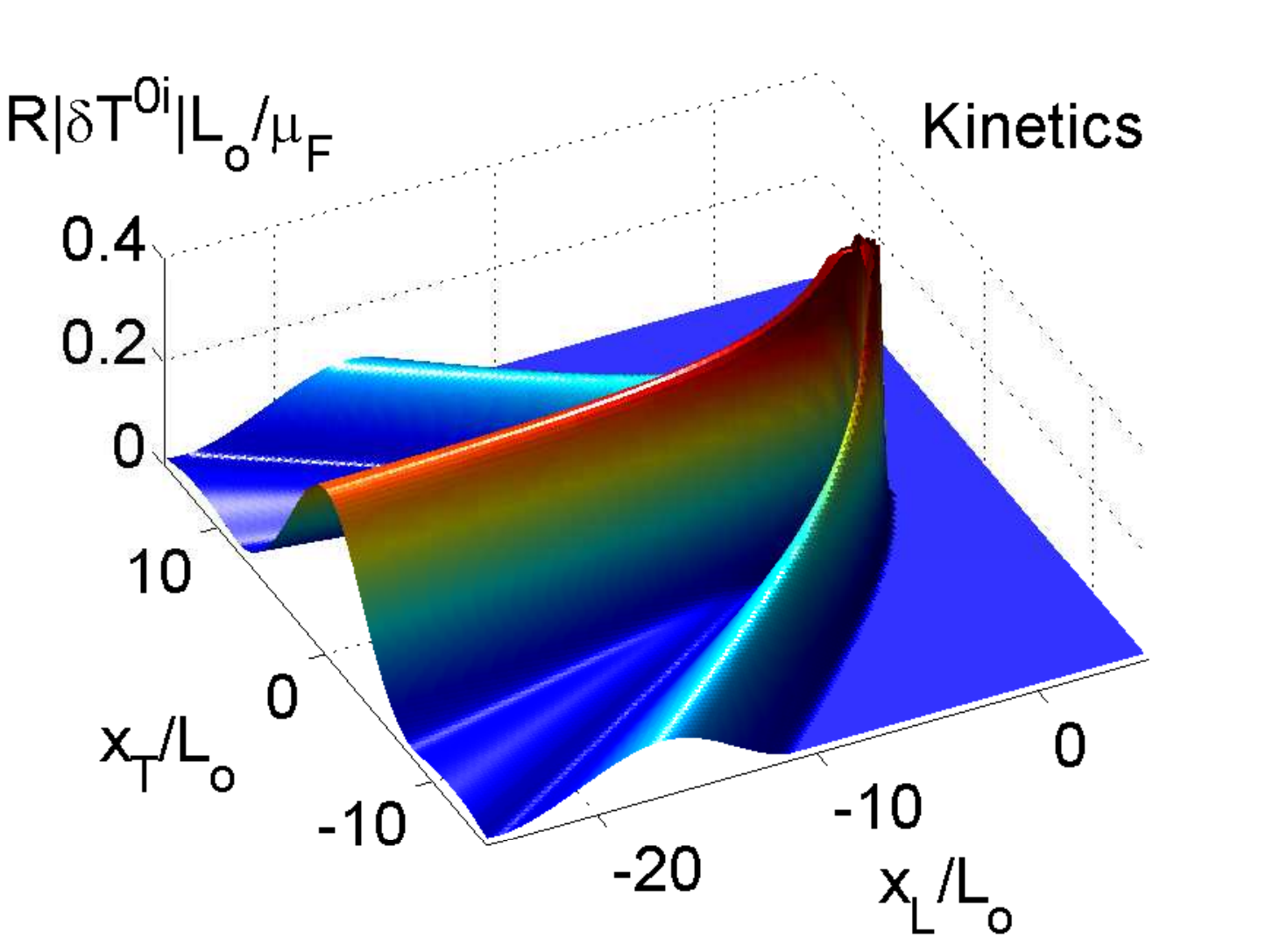}
\includegraphics[width=0.7\textwidth]{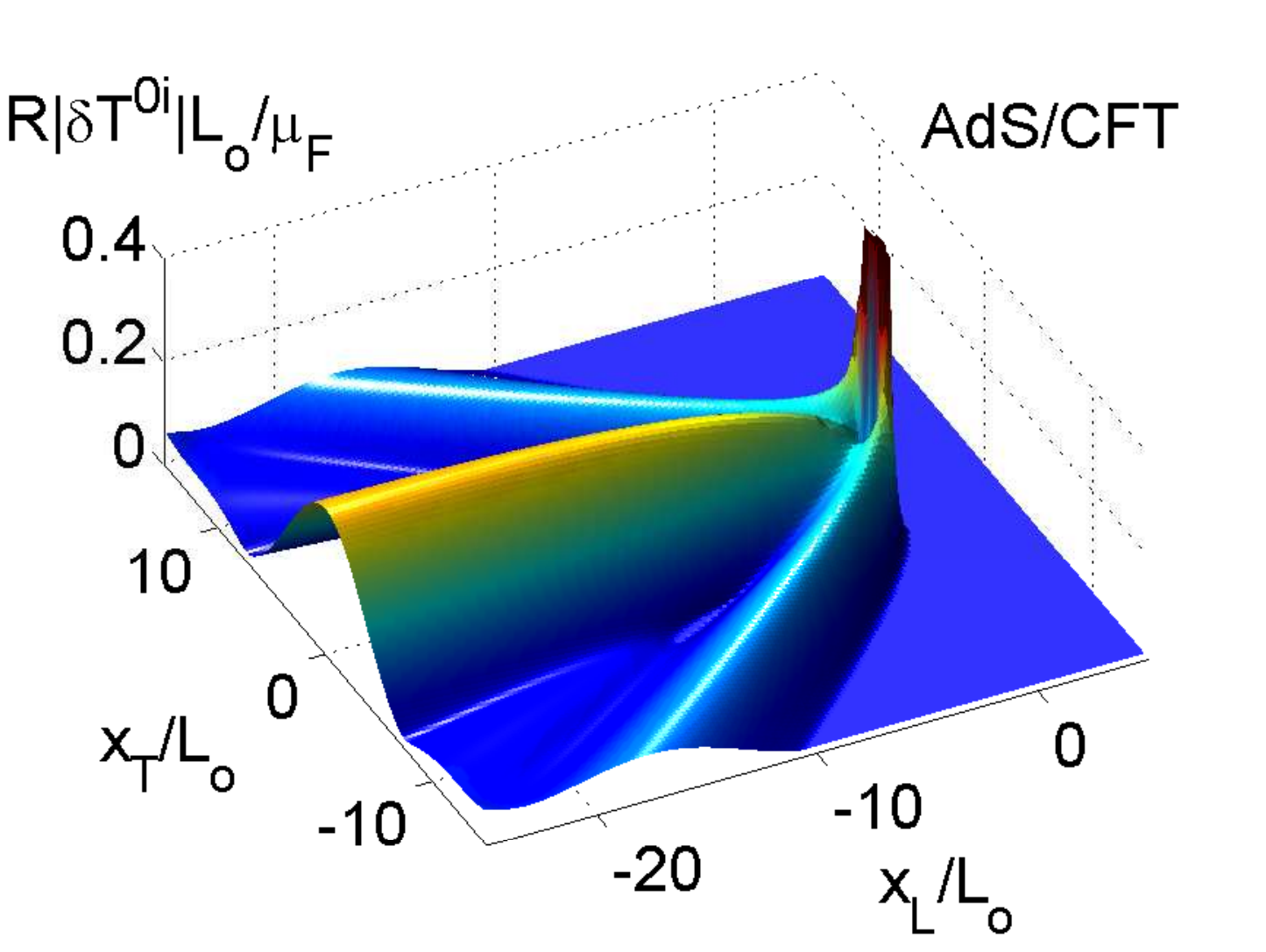}
\end{center}
\caption{
The magnitude of the Poynting vector $|T^{0i}|$ (in scaled units) times $R= \sqrt{x_T^2 + x_L^2}$ that is induced
by a heavy quark probe in (a) weakly coupled QCD and (b) strongly coupled
$\N=4$ SYM. 
Here $\Lo$ is the shear length and the $\mu_F(v)$ is the 
drag coefficient for each case (see text).
\label{mEPfull}
}
\end{figure}

Using the formalism outlined in the previous section we compute
the energy density and Poynting vector induced by the heavy quark in
both kinetic theory and the AdS/CFT correspondence.
To compare the AdS/CFT and the kinetic 
theory results  we have measured all length-scales  in units 
of the shear length
\st
\label{Ldef}
    \Lo \equiv  \frac{\textstyle{\frac{4}{3} } \eta c }{(e + \Pr)c_s^2} 
\stp
where $c_s^2$ is the squared sound speed
and in practice the speed of light is set to unity.
$\Lo$ is proportional to the mean free path in kinetic theory
and equal to $1/\pi T$ for the $\N=4$ theory.  
 At large distances, where ideal
hydrodynamics is applicable, the amplitude of the disturbance is proportional to
the strength of the energy loss. Thus,  we divide the response by the corresponding drag coefficient $\mu_F(v)$ for each theory, \Eq{mudef_kt} and \Eq{mudef_ads}. 
With these rescalings the 
two theories produce the same (rescaled) stress tensor at asymptotically large distances,  but differ in their approach to the ideal hydrodynamic limit
as we will analyze in \Sect{hydro:sec}. \Fig{EPfull} and \Fig{mEPfull}  
compare the non-equilibrium stress 
in the two cases.  A complete discussion is reserved for the summary in  \Sect{discussion}.  


\begin{figure}
\begin{center}
\includegraphics[width=0.49\textwidth]{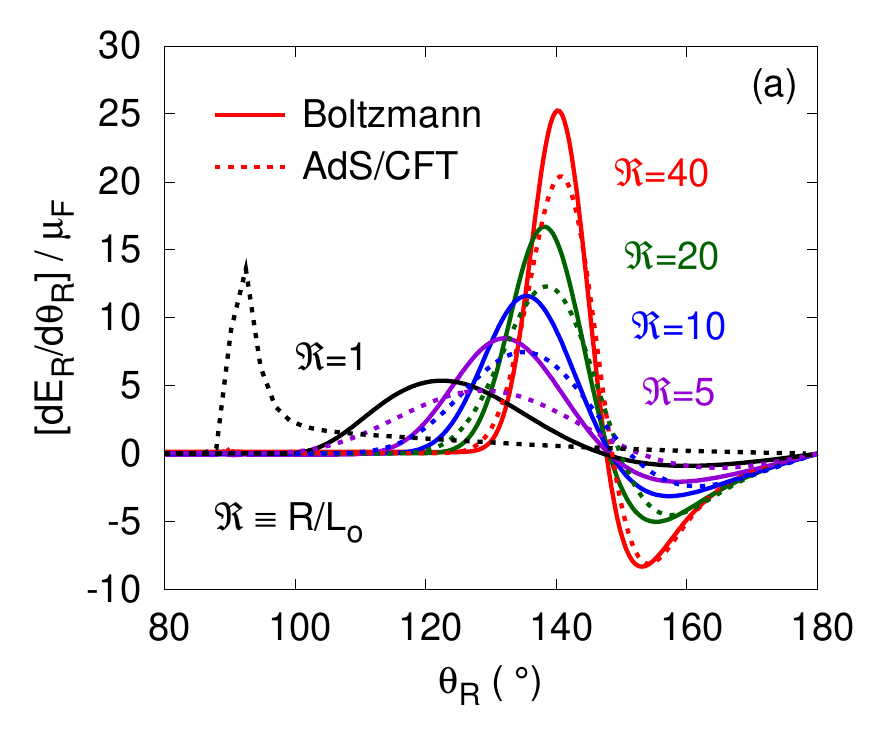}
\includegraphics[width=0.49\textwidth]{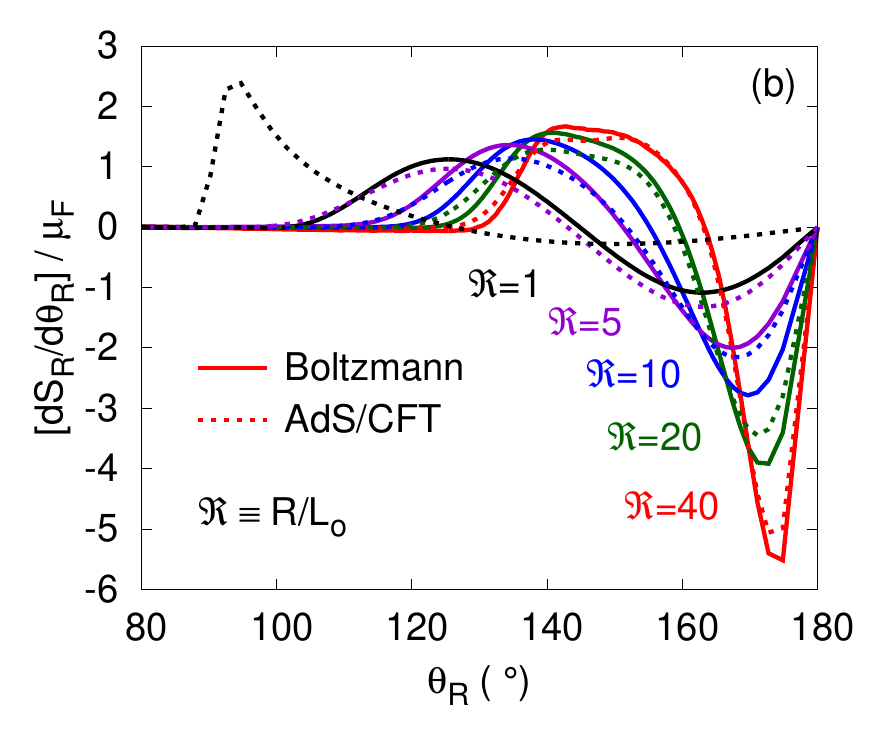}
\caption{The angular distribution of (a) the energy density 
$[\rm{d}E_R/\rm{d}\theta_R]/\mu_F$ and 
(b) the energy flux $[\rm{d}S_R/\rm{d}\theta_R]/\mu_F$ 
at distances $\RR=1, \, 5, \, 10, \, 20, \, \mbox{and}\, 40$ 
for the kinetic theory and gauge gravity duality.
Here $\Lo$ is the shear length and the $\mu_F(v)$ is the 
drag coefficient for each case (see text).
\label{EPcompare}}
\end{center}
\end{figure}

In order to compare the stress tensor quantitatively
we plot the energy density  in concentric 
circles of radius $R$ around the head of the quark. 
Specifically, we define 
\st
\label{dedrdef}
\frac{\rm{d}E_R}{\rm{d}\theta_R}
=2\pi R^2\sin\theta_R \, \delta T^{00} (\mathbf{R}) \, ,
\stp
where $\mathbf{R}=x_T\hat{\x}_T+x_L\hat{\z}$ and the polar angle
is measured from the direction of the quark $\hat{\z}$:
\st
\label{thetardef}
\begin{minipage}[c]{0.14\textwidth}
\includegraphics[width=\textwidth]{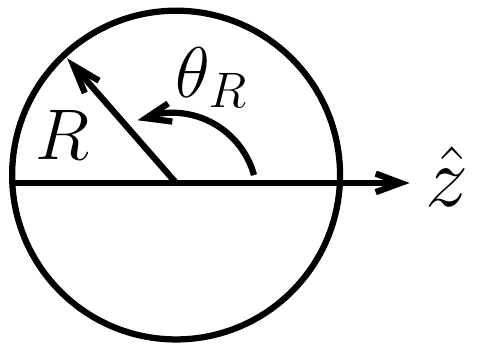} 
\end{minipage} \, .
\stp
Similarly, the angular distribution of the energy flux is given by  
\bg
\label{dsdrdef}
\frac{\rm{d}S_R}{\rm{d}\theta_R}
&=&2\pi R^2 \sin\theta_R \, \hat{R}^i \delta T^{0i}(\mathbf{R}) \, ,\nonumber\\
&=&2\pi R^2 \sin\theta_R \left[\cos\theta_R \, \delta T^{0z} (\mathbf{R})+
\sin\theta_R \, \delta T^{0x} (\mathbf{R})\right] \, .
\nd
Numerical results for the angular distributions of the energy 
density and flux at several scaled distances  $\RR \equiv R/\Lo$
are shown in \Fig{EPcompare}. 

%


There is a dramatic change  in the AdS/CFT curves between $\RR=1$ and $\RR=5$,
indicating a transition from hydrodynamic behavior to
quantum dynamics at distances of order $\sim 1/\pi T$.
Since  this quantum dynamics 
lies beyond the semi-classical Boltzmann approximation, no transition
is seen in the kinetic theory curves.
 It would be interesting to calculate
the stress tensor in this region perturbatively to better understand
the differences between   the two theories for $R \sim 1/\pi T$. 

Let us pause to discuss the limitations of both calculations.  The point of the
current work is to compare the approach to hydrodynamics at infinitely weak and
infinitely strong coupling.  In both cases the coupling is taken to zero or
infinity before the limit, $v\rightarrow1$.  As we now discuss, this limits the
lengths scales that can be meaningfully studied in both theories.

In the kinetic theory calculation the resulting stress tensor is valid for
distances, $R \gg 1/(g^2 T \log g^{-1})$. For distances shorter than $1/(g^2
T\log g^{-1})$ the collisionless non-abelian Vlasov equations should be used to
describe the medium response at weak coupling
\cite{Elze:1989un,Blaizot:2001nr}.
However, for distances longer than $1/(g^2 T \log g^{-1})$ the effect of the
plasma dynamics is incorporated into the polarization tensor of the soft
collisional integrals between the heavy quark and the particles
that make up the bath. For example, the drag coefficient computed
by Braaten and Thoma includes an HTL propagator in the t-channel exchange that
includes the plasma physics of the polarization tensor
\cite{LeBellac}. We have limited the
evaluation of this polarization tensor to a leading log approximation.

Further, the weakly coupled calculation is limited to modest $\gamma$.  We
have implicitly taken the coupling constant to zero  before taking
$v\rightarrow 1$ so that the radiative energy loss of the heavy quark can be
neglected.  For a  small but finite coupling constant  radiative energy loss is
suppressed when the Lorentz gamma factor of the heavy quark is not  too large
\cite{Moore:2004tg}, $\gamma  \lsim \frac{m_D}{T\alpha_s} \sim 1/g$.  
  
Similarly, the AdS/CFT calculation is limited to comparatively large distances
$x_{L}, x_{T} \gg 1/\sqrt{\gamma} \pi T$.   In \Sect{adscft} we introduced
a new set of helicity variables so that the medium response on distances
much greater than $1/\sqrt{\gamma} \pi T$ can be determined in 
the limit when  $v=1$.
For distances much less than $1/\sqrt{\gamma} \pi T$
 the structure of stress tensor  
 has been analyzed in detail \cite{Yarom:2007ap,Gubser:2007nd,Noronha:2008un}.
The medium response is characterized by a transverse length scale of $1/\sqrt{\gamma} \pi T$,
and a corresponding  longitudinal scale of $1/\gamma^{3/2} \pi T$, where
$\gamma$ is the Lorentz factor of the heavy quark. 
In \Fig{EPcompare} we are investigating distances  of order
$x_{L}, x_T \sim 1/\pi T$  and the physics associated with these very short scales is not visible.
Although the importance of the $1/\sqrt{\gamma}\pi T$ scale was understood in the context of momentum fluctuations \cite{Gubser:2006nz,CasalderreySolana:2007qw,Dominguez:2008vd}, 
it is instructive to see these scales reappear in the asymptotic expansion for the induced stress tensor
at short distances \cite{Yarom:2007ap,Gubser:2007nd}.  Rewriting Eq.~(137) of \Ref{Gubser:2007nd}  
 in terms of $\gamma$,  and expanding
for $\gamma$ large  with $\tilde x_{L} \equiv \gamma x_L$ of order $ x_T$, we have
\st
\label{adscfteq_short}
\frac{1}{\sqrt{\lambda} }\llangle \delta T^{00}(x_L, x_T) \rrangle = 
 \frac{\gamma^2 x_T^2}{6 \pi ^2 \left( \tilde x_L^2+x_T^2\right)^3} +  \frac{T^2\gamma^3}{24}  \frac{ \left(2 \tilde x_L^3+ \tilde x_L x_T^2\right)}{ \left(\tilde x_L^2+x_T^2\right)^{5/2}} +  \ldots \, .
\stp
The first term is the leading term at short distances and is independent of temperature.
The second term (which captures the first finite temperature correction) is
subleading in inverse powers of distance, but enhanced by a power of $\gamma$.
Comparing the
magnitude of these terms we see that the first term will dominate
provided 
\st
   \sqrt{(\gamma x_L)^2 + x_T^2} \lsim \frac{1}{\sqrt{\gamma} \pi T}  \, .
\stp
This constraint  limits the validity (and utility) of the short distance expansion to rather short distances. 

In summary we are examining two extreme limits -- infinitely weak and infinitely
strong coupling.  The disadvantage of this approach is that some of the marked
differences  at short distances between the Vlasov response of weakly
coupled QCD and the AdS/CFT response are not visible (see especially \cite{Betz:2008wy}).  The advantage of this
approach is that the onset of hydrodynamics can be clearly compared. 
We will analyze the hydrodynamic limit in the next section.

\section{Hydrodynamic Analysis}
\label{hydro:sec}

At large distances (see \Fig{EPfull} and \Fig{mEPfull}), the medium response to the heavy quark probe 
clearly exhibits hydrodynamic flow. 
Following in part the discussion by Chesler
and Yaffe \cite{Chesler:2007sv},
we will analyze this
hydrodynamic response order by order in the gradient expansion for kinetic
theory and for  the AdS/CFT correspondence. 
The strongly coupled $\N=4$ theory is conformal and the appropriate hydrodynamic theory is conformal hydrodynamics \cite{Baier:2007ix,York:2008rr}. 
Similarly, to leading order in the coupling constant QCD is also conformal  and again
conformal hydrodynamics is applicable  in this limit.
 Beyond leading order, there are corrections to 
kinetic theory which break scale invariance and non-conformal
hydrodynamics must be used to characterize the long wavelength response 
\cite{Romatschke:2009kr,Hong:2010at}.

For both kinetic theory and the AdS/CFT, the stress tensor of the full theory 
satisfies the conservation law  
\Eq{micro-eom} 
At large distances the stress tensor is described by hydrodynamics
up to uniformly small corrections suppressed by inverse powers 
of the distance.
In hydrodynamics, the spatial components of the stress tensor are specified by 
the constituent relation order by order in  the gradient expansion.
Specifically, the stress tensor  can be written
\begin{equation}
T_{\rm hydro}^{\mu\nu}=(e+\mathcal{P})u^\mu u^\nu +\mathcal{P}g^{\mu\nu}
+\pi^{\mu\nu} \, ,
\end{equation}
where the dissipative part of the stress tensor $\pi^{\mu\nu}$ 
is expanded in gradients of $T^{00}$ and $T^{0i}$ or ($T$ and $u^{\mu}$) 
to a specified order. For linearized conformal hydrodynamics (where $u^{\mu}=(1,{\bm u} )$)
 this expansion through second order in gradients reads \cite{Baier:2007ix}
\begin{align}
\label{pimunu_stat}
\pi^{ij}=& -2\eta\llangle \partial^{i} u^{j} \rrangle  -2\eta\tau_\pi \llangle \partial^i\partial^j\ln T \rrangle  \qquad \qquad \mbox{(Static)} \,  ,
\end{align}
and all temporal components are zero.
Here $\langle \, \cdots \, \rangle$ denotes the symmetric and traceless
spatial component of the bracketed tensor \cite{Baier:2007ix}, \emph{i.e.} 
for linearized hydrodynamics we have
\st
\llangle \partial^{i} \partial^j  \ln T \rrangle = \left(\partial^{i} \partial^j - \frac{1}{3} \delta^{ij} \partial^2 \right) \ln T \, .
\stp
We will refer to the conservation laws 
together with the constituent
relation (\Eq{pimunu_stat}) as the  static form of second order hydrodynamics.
Using the  lowest order equations of motion (ideal hydrodynamics) and conformal symmetry, the second order term $\llangle \partial^{\mu} \partial^\nu \ln T\rrangle $ can be replaced by the time derivative of $\pi^{\mu\nu}$ \cite{Baier:2007ix}
\begin{equation}
\pi^{ij}=
-2\eta\langle \partial^i u^j \rangle -\tau_\pi\partial_t
\pi^{ij} \, \qquad \qquad  \mbox{(Dynamic)} \,  .
\end{equation}
This rewrite of the constituent relation can be interpreted as a dynamical
equation for $\pi^{\mu\nu}$, and is  similar to the second order form
of Israel and Stewart \cite{Israel:1976tn,Israel:1979wp}. We will refer to this equation of motion for $\pi^{\mu\nu}$, together with the conservation laws  as the dynamic form of second order hydrodynamics.  
 
At long distances, the  form  of the stress-energy tensor is described by 
$T_{\rm hydro}^{\mu\nu}$ up to terms suppressed by 
inverse powers of the  distance.  We will express the full 
stress tensor  as a hydrodynamic term,  which is irregular in the limit of
$\omega, \k \rightarrow 0$, plus a correction which we will verify is
a regular function for $\omega,\k\rightarrow 0$
\begin{equation}
T^{ij}=T_{\rm hydro}^{ij}\left[T^{00},T^{0i}\right]+\tau^{ij} \, .
\end{equation} 
We have temporarily emphasized here that $T_{\rm hydro}^{ij}$ is a 
functional of the densities $T^{00}, T^{0i}$ as specified 
by the constituent relation and the equation of state.
Then the equation of motion in Fourier space becomes
\begin{equation}
\label{hydro_with_source}
-i\omega \, \delta T^{0j}+ik^i \, \delta T_{\rm hydro}^{ij}
= S^{j}_{\rm hydro} (\omega,\k) \, ,  
\end{equation}
where
\st
S^{j}_{\rm hydro}(\omega,\k) \equiv \frac{\dd p^j}{\dd t} 2\pi \delta(\omega - \v \cdot \k)  - ik^i\tau^{ij}  \, .
\stp

Examining $S_{\rm hydro}$, we see that $-ik^{i} \tau^{ij} $ acts as an additional source term for
hydrodynamics. What makes this decomposition useful is  that $\tau^{ij}$  (in contrast to $T_{\rm hydro}^{ij}$) is
a regular function  at small $\k$ and $\omega$.
For the steady state problem we are considering, $\tau^{ij}$ can be written with three functions proportional to 
the symmetric tensors consisting of $\v$ and $\k$ 
\begin{multline}
\label{tauijbasisfcns}
\tau^{ij}(\omega,k^2)  \equiv 
2\pi\mu_F\delta(\omega- \v\cdot\k) \,
\left[ \, \left(v^iv^j-\third v^2\delta^{ij}\right) 
\phi_1(\omega,k^2) \right. \\
+\left. \left(iv^i k^j+ik^{i}v^{j}-i\twothirds v_l k^l
\delta^{ij}\right)  \phi_2(\omega,k^2)
+\left(k^ik^j-\third k^2\delta^{ij}\right)  
\phi_3(\omega,k^2) \, \right] \, , 
\end{multline}
where $\phi_1,\phi_2$ and $\phi_3$ are regular for $\k\rightarrow 0$. 
The source can be expressed similarly 
\st
\label{hydrosrcfcns}
{\bf S}_{\rm hydro} \equiv 2\pi \delta(\omega - \v\cdot \k) \left[ \phi_v(\omega,k^2) \v +  \phi_k(\omega,k^2) i\k \right] \, .
\stp
Since $\tau^{ij}$ is localized, 
we can expand it for small $\omega$ and  $\k$. 
Using  an obvious  notation for the Taylor series,
\st
 \phi_{1}(\omega, k^2) \simeq \phi_{1}^{(0,0)} + \phi_{1}^{(1,0) } (-i\omega) + \frac{1}{2!} \left[ \phi_{1}^{(2,0)}\, (-i\omega)^2 +  \phi_1^{(0,2)} \,  (ik)^2  \right]  + O(k^3) \, ,
\stp
we see that the full source for hydro through second order can be expressed   
in terms of three expansion coefficients, $\phi_1^{(0,0)}$ and  $\phi_1^{(1,0)}$, and $\phi_{2}^{(0,0) }$:
\begin{multline}
\label{Shydroexpand}
{\bm S}_{\rm hydro} = 2\pi \mu_F \delta(\omega - \v\cdot \k) 
\Big[ \underbrace{\left(1 - i\omega \phi_1^{(0,0)} - \phi_{1}^{(1,0)} \omega^2 + \phi_2^{(0,0)}k^2 \right)}_{\equiv \phi_v} \v  \\
+ \underbrace{ \left( \third v^2 \phi_{1}^{(0,0)} - \third v^2 \phi_{1}^{(1,0)} i\omega  - \third\phi_2^{(0,0)} i\omega  \right) }_{\equiv \phi_k} i\k\Big] + O(k^3) \, .
\end{multline}

Summarizing, $\tau^{ij}$  can be determined by comparing 
the full numerical solution for $T^{ij}$  to
$T_{\rm hydro}^{ij}$.
Then, by fitting the functional
form given by an expanded \Eq{tauijbasisfcns},  we can extract the three coefficients 
$\phi_1^{(0,0)}, \, \phi_1^{(1,0)}$, and $\phi^{(0,0)}_2$ 
for the Boltzmann equation and the AdS/CFT correspondence. These 
coefficients fully
specify the hydrodynamic source of a heavy quark through quadratic order.
\app{source_details} gives some sample fits to our numerical results 
and the fit coefficients are collected in ~Table~\ref{sourcecoeff}.
The quality of the fits  given in \app{source_details} indicates  that
$\tau^{ij}$ is well described by a polynomial at small $k$ and $\omega$ 
and justifies the analysis of this section.

Examining Table ~\ref{sourcecoeff}, we notice that in the Boltzmann case the
expansion coefficients proportional to $\phi_1$ vanish.  In fact,
$\phi_1(\omega,k^2)$ vanishes   to all orders in $\omega,k$. 
This follows from
rotational symmetry around the $\k$ axis and the somewhat
special form of the kinetic theory source in \Eq{source}. Since we do not expect
this property to hold beyond the leading log approximation, further discussion
of this point is relegated to \app{source_details}. 




\begin{table*}[t]
\begin{center}
\begin{tabular}{|l||c||c|c|}
\hline & \, $\phi_1^{(0,0)}$ /$\Lo$ \, & 
\, $\phi_1^{(1,0)}/\Lo^2$ \, & 
\, $\phi_2^{(0,0)}/\Lo^2$ \,  \\
\hline 
\, Boltzmann \, & 0 & 0 & 0.484 \\ 
\hline 
\, AdS/CFT \, & -1 & -0.34 & -0.33 \\
\hline
\end{tabular}
\label{sourcecoeff}
\end{center}
\caption{Table of hydrodynamic source coefficients. The equations 
of motion are given by second order hydrodynamics with 
a source term, \Eq{hydro_with_source}. The  source term is expanded 
to quadratic order in $\k$ and $\omega$  in \Eq{Shydroexpand} which
defines these coefficients.
The first coefficient
$\phi_{1}^{(0,0)}$ was computed analytically in the AdS/CFT case
by Chesler and Yaffe \cite{Chesler:2007sv}. 
Here  $\Lo$ is the shear length (see text). 
}
\end{table*}

With the source functions $\phi_v(\omega,k^2)$ and $\phi_k(\omega,k^2)$ 
known (numerically) through quadratic order, the hydrodynamic
approximation to the equations  of motion  
for (static) second order hydrodynamics 
reads  (with $v=1$)
\begin{align}
-i\omega \, \delta T^{0z'}+  c^2(k) \, ik\delta T^{00}
+\Gamma_sk^2 \, \delta T^{0z'}=&
\left[\cos\theta \, \phi_v+ik \, \phi_k \right]  
2\pi\mu_F\delta(\omega-\v\cdot\k) \, , \nonumber\\
-i\omega \, \delta T^{0x'}+Dk^2 \, \delta T^{0x'}
=&\sin\theta \, \phi_v \, 
2\pi\mu_F\delta(\omega-\v\cdot\k) \, ,
\end{align}
where $z'$ points along the $\k$ axis and  $x'$ is perpendicular to $\k$ 
(see \app{details}). In these  equations
$\Gamma_s=(4\eta/3)/(e+\mathcal{P})$, $D=\eta/(e+\mathcal{P})$. 
and $c^2(k) = c_s^2(1 + \tau_\pi \Gamma_sk^2)$. 
Using these approximate expressions  and the exact equation
\st
-i\omega \, \delta T^{00}+ik \, \delta T^{0z'}=2\pi\mu_F \, 
\delta(\omega-\v\cdot\k) \, ,\\
\stp
the hydrodynamic solutions in Fourier space are given by 
\begin{subequations}
\label{hydrosol}
\begin{eqnarray}
\delta T^{00}(\omega,\mathbf{k})
&=&\frac{i\left[\omega+k\cos\theta\, \phi_v\right]
-k^2\left[\Gamma_s+\phi_k\right]}
{\omega^2-c^2(k)\, k^2+i\Gamma_s \, \omega k^2}
\, 2\pi \mu_F \delta(\omega-k\cos\theta) \, , \\
\delta T^{0x'}(\omega,\mathbf{k})
&=&\frac{i\sin\theta\, \phi_v}
{\omega+iD(\omega) \, k^2} \, 
2\pi \mu_F \delta(\omega-k\cos\theta) \, .\\
\delta T^{0z'}(\omega,\mathbf{k})
&=&\frac{i\left[ \, \omega\cos\theta\, \phi_v 
+c^2(k) \, k \, \right]
-k\omega \, \phi_k}
{\omega^2-c^2(k) \, k^2+i\Gamma_s \, \omega k^2}\,
 2\pi \mu_F \delta(\omega-k\cos\theta) \, .
\end{eqnarray}
\end{subequations}
The solutions  can also be used  for
first order hydrodynamics  provided the wave speed $c^2(k)$ and the
source functions $\phi_v$ and $\phi_k$ 
are truncated at leading order, {\it i.e.}  $c^2(k) \rightarrow c_s^2$ and
$\phi_1(\omega,k^2) \approx \phi_1^{(0,0)}$. 
Similarly, the hydrodynamic solutions for the dynamic implementation of second order
hydrodynamics takes the same functional form as \Eq{hydrosol}
with the replacements
\st
  c^2(k) \rightarrow c_s^2 \, ,   \qquad  \Gamma_s \rightarrow \Gamma_s(\omega) \equiv \frac{\Gamma_s}{1-i\tau_\pi\omega}\, ,  \qquad  D \rightarrow D(\omega) \equiv \frac{D}{1 -i\tau_\pi \omega} \, .
\stp

\begin{figure}
\centering
\includegraphics[width=80mm]{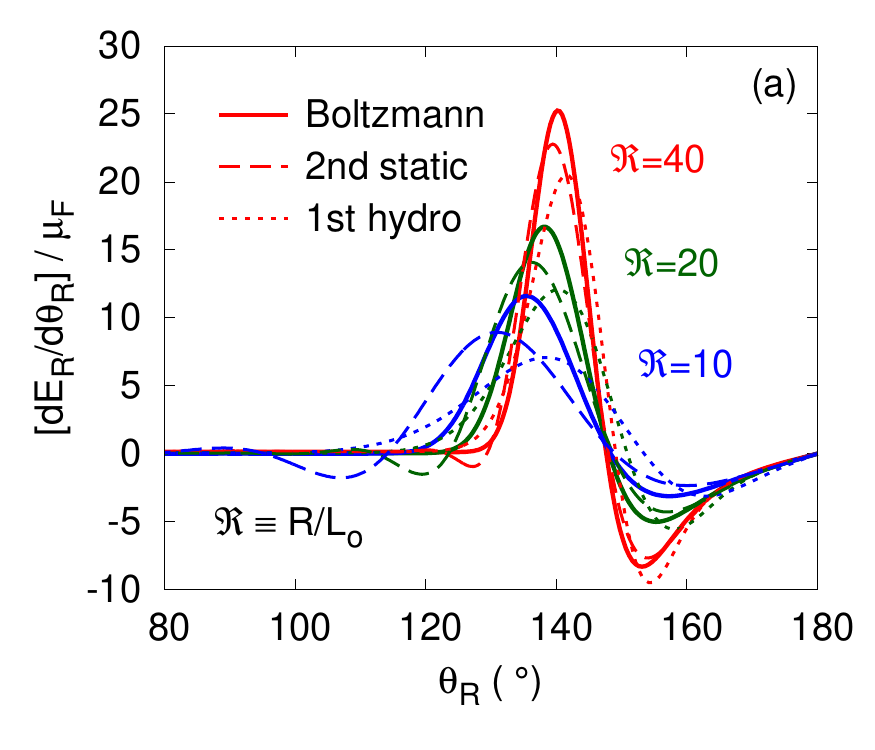}
\includegraphics[width=80mm]{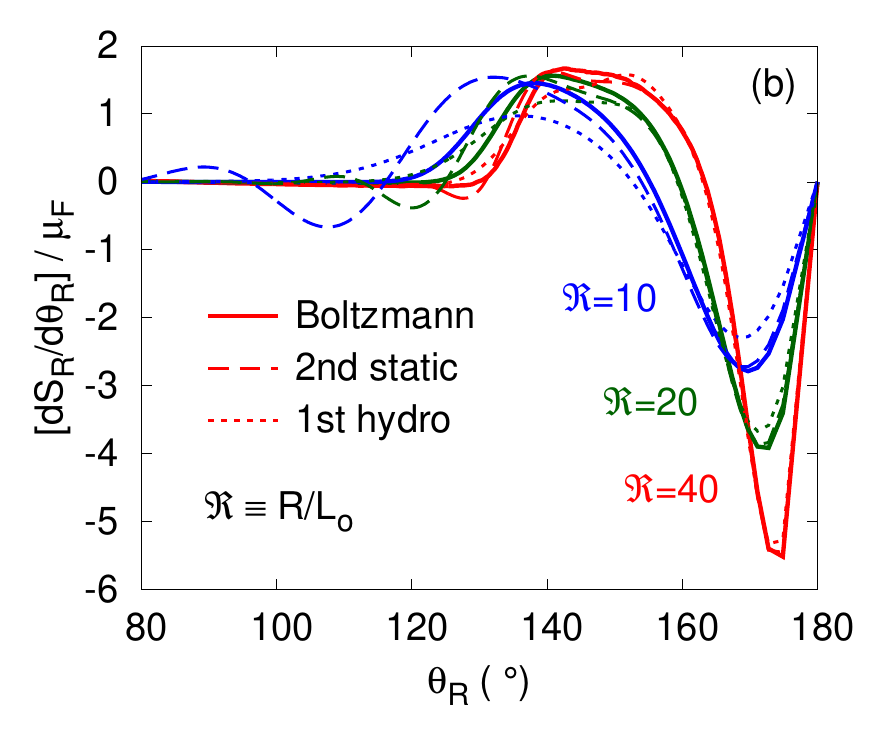}
\caption{The angular distribution of (a) the energy density 
$[\rm{d}E_R/\rm{d}\theta_R]/\mu_F$ and   
(b) the energy flux $[\rm{d}S_R/\rm{d}\theta_R]/\mu_F$
given by the Boltzmann equation at distances 
$\RR=10, \, 20, \, \mbox{and} \, 40$. 
The Boltzmann results are compared with the first order and second order static hydrodynamics. 
}
\label{Boltzh}
\end{figure}

\begin{figure}
\centering
\includegraphics[width=80mm]{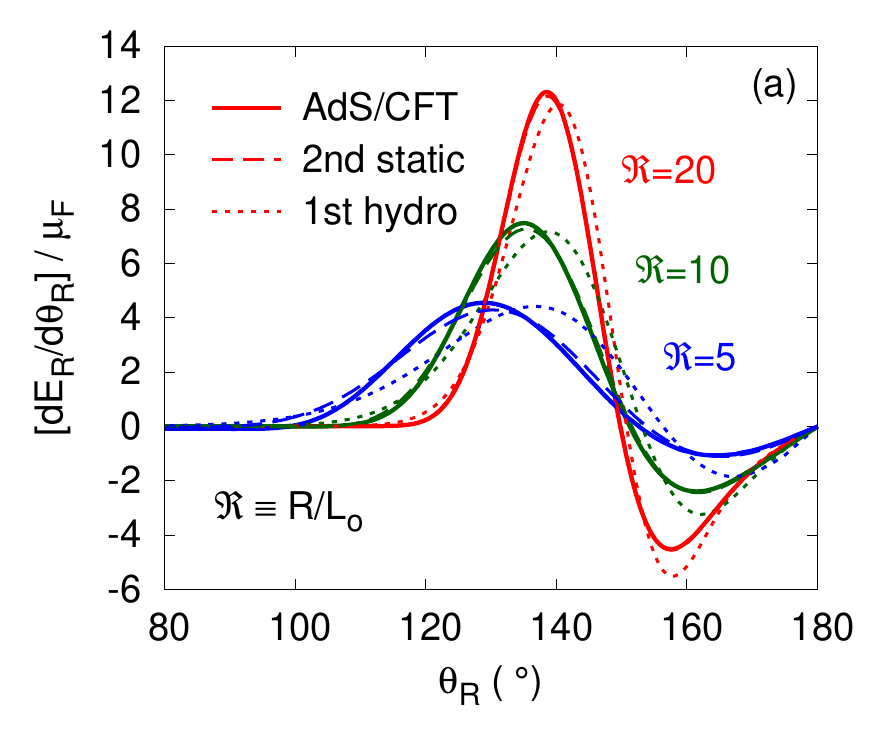}
\includegraphics[width=80mm]{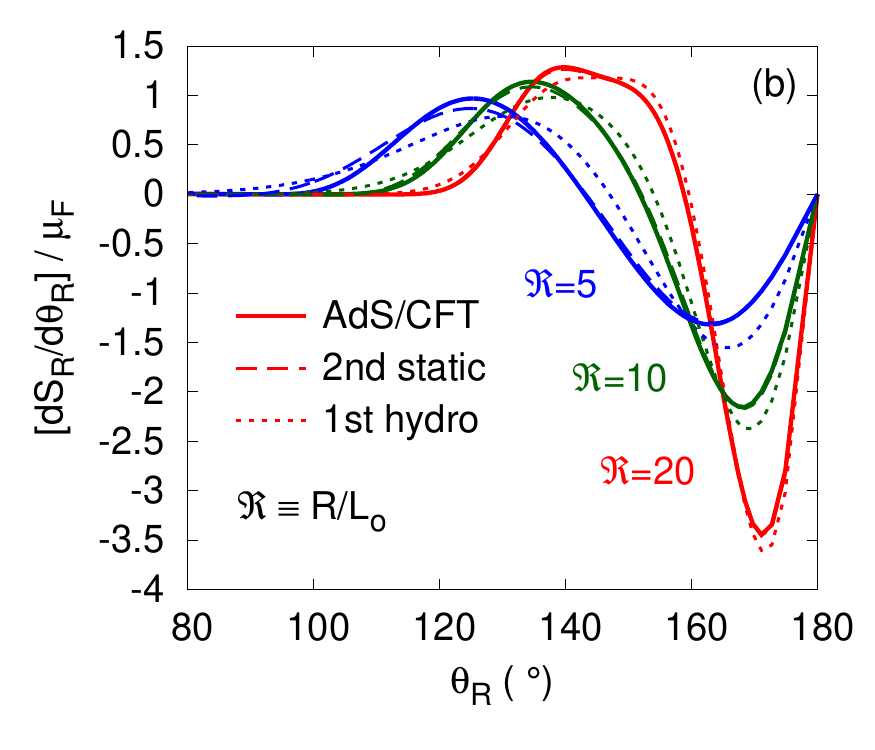}
\caption{The angular distribution of (a) the energy density 
$[\dd E_R/\dd \theta_R]/\mu_F(v)$ 
and (b) the energy flux 
$[\dd S_R/\dd \theta_R]/\mu_F(v)$
given by the AdS/CFT correspondence at distances 
$\RR=5, \, 10, \, \mbox{and} \, 20$. 
The AdS/CFT results are compared with the first order and second order static hydrodynamics.   Here $\Lo=1/\pi T$ is the shear length  and $\mu_F(v) = \gamma \sqrt{\lambda} \pi T^2/2$ is the heavy quark drag coefficient for the AdS/CFT.
\label{AdSh}
}
\end{figure}

\begin{figure}
\begin{center}
\includegraphics[width=0.49\textwidth]{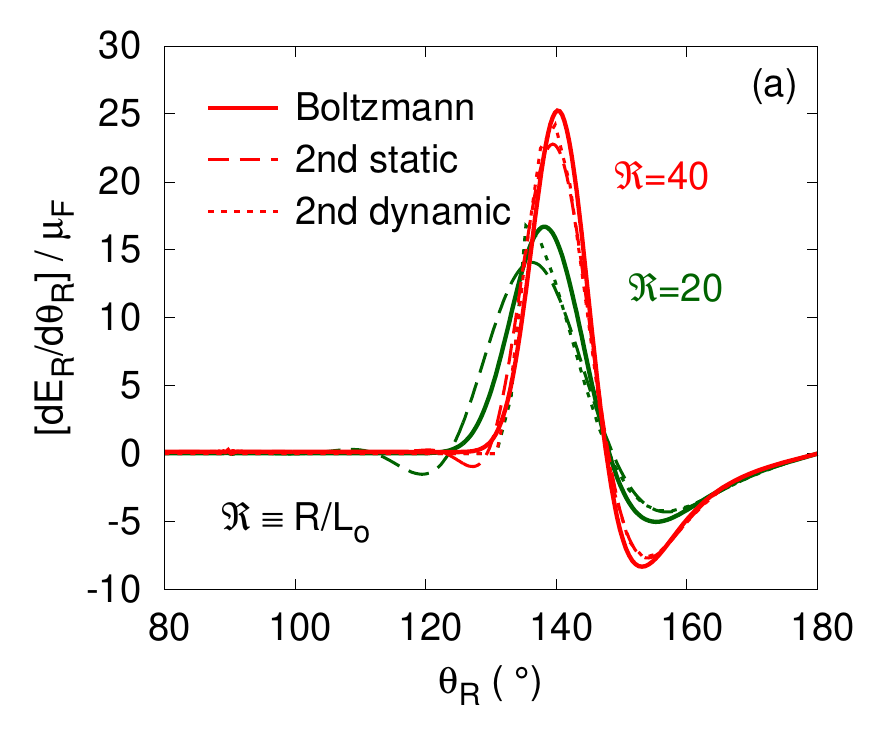}
\includegraphics[width=0.49\textwidth]{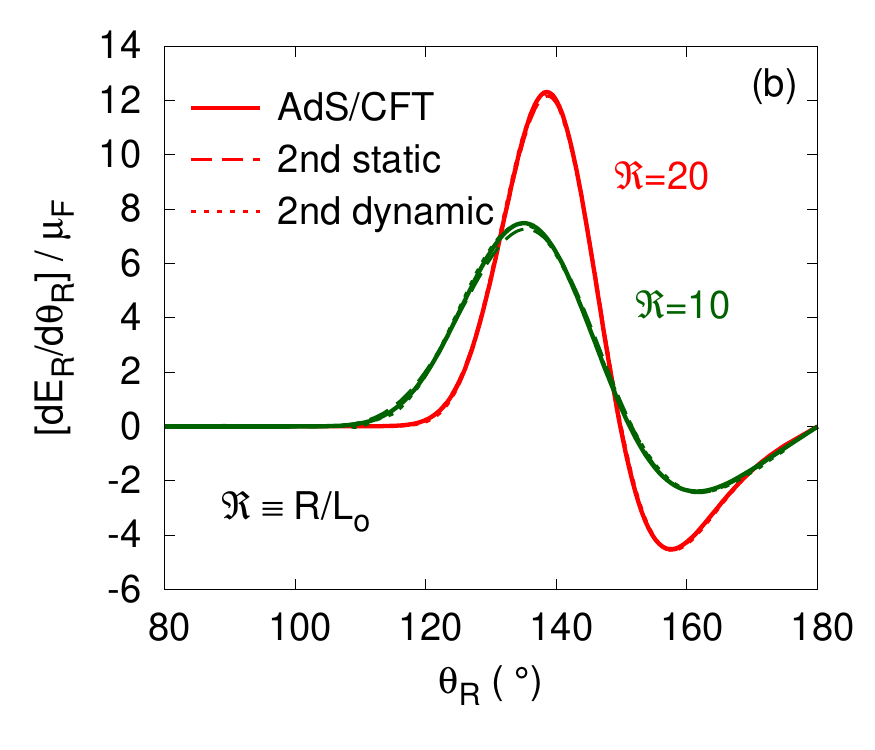}
\end{center}
\caption{The angular distribution of  the energy density 
$[\rm{d}E_R/\rm{d}\theta_R]/\mu_F$ 
in (a) kinetic theory and (b) the AdS/CFT.
The kinetic theory curves are plotted at distances $\RR= 20, \, 40$  
while the AdS/CFT curves are plotted at distances $\RR= 10, \, 20$. 
The Boltzmann and AdS/CFT results are compared to the static and dynamic
implementations of second order  hydrodynamics (see text). The 
differences  between the static and dynamic implementations of 
second order hydrodynamics reflects the size of neglected third order
terms. Here $\Lo$ is the shear length and $\mu_F(v)$ is the drag coefficient
for each theory.
\label{2ndhydro1}
}
\end{figure}

Given these hydrodynamic solutions and the hydrodynamic source functions tabulated 
in Table~\ref{sourcecoeff},
 the hydrodynamic stress tensor in coordinate space can be computed using numerical Fourier transforms. The stress tensor of
first and second order hydrodynamics (with the corresponding source) 
is compared to the full kinetic theory stress tensor in \Fig{Boltzh}. 
\Fig{AdSh} presents the analogous AdS/CFT results.  Finally,
a comparison between the static and dynamic implementations of second
order hydrodynamics  is given in  ~Figure~\ref{2ndhydro1} and
provides an estimate of higher order terms in the hydrodynamic expansion.  We will discuss these results in the next section.



\section{Summary and Discussion}
\label{discussion}

To keep this discussion self contained, we first recapitulate the problem and
the  corresponding notation. An infinitely heavy quark moves along the $z$ axis
with velocity $v\simeq c$, depositing energy  and disturbing  the surrounding
equilibrium  plasma.  
We presented and compared the energy and momentum density distributions 
in two distinctly different plasmas -- a weakly coupled QCD plasma described by kinetic  theory and a strongly coupled $\N=4$ plasma described by the  AdS/CFT correspondence.
The steady state  stress tensor distributions can
be written in cylindrical coordinates  and comoving coordinates (see
Eqs~\ref{cylindrical1} and \ref{cylindrical2}).  
\Fig{EPfull} and \Fig{mEPfull} exhibit 
the energy density $T^{00}$ and the magnitude of the Poynting 
vector $|{\bf S}| \equiv |T^{0i}|$ for kinetic theory and the $\N=4$ theory
respectively. 
To compare these theories we measured all distances
in terms of a length scale given by a combination of hydrodynamic parameters 
\[
    \Lo \equiv  \frac{\textstyle{\frac{4}{3} } \eta c }{(e + \Pr)c_s^2} \, .
\]
This length  is of order the mean free path in kinetic theory and equals $1/\pi T$ 
for the AdS/CFT.
In each theory we divided the stress tensor by the corresponding heavy
quark drag coefficient $\mu_F(v)$ so that at asymptotically large distances (where ideal hydrodynamics is valid) the rescaled stress tensors  are equal.  
At asymptotic distances both theories reproduce the ``Mach cone'' structure
characteristic of ideal hydrodynamics, but these model plasmas differ 
at sub-asymptotic distances in their approach to this ideal hydrodynamic regime.
In particular  the Boltzmann theory is considerably 
{\it less} diffuse  than   the AdS/CFT. 
In kinetic theory the short distance 
response is  reactive, and   the sharp band seen in \Fig{mEPfull}(a)  (which is indicative of free streaming quasi-particles) is absent in \Fig{mEPfull}(b).
 We will see that 
the response of the AdS/CFT closely follows the predictions of
hydrodynamics at modest distances which is strongly damped by the shear viscosity at short distances. 
This difference between the two can also be seen quantitatively in \Fig{EPcompare}
which compares the kinetic theory  and AdS/CFT results by plotting the
energy density and energy flux at concentric  circles of radius $R$ in scaled
units: 
\[ 
\RR = R/\Lo \, .
 \] 
The precise definitions of
$dE_R/d\theta_R$ and $\dd S_R/\dd\theta_R$ is given by Eqs.~\ref{dedrdef}, \ref{thetardef},
and \ref{dsdrdef}  respectively. As  discussed in
the previous paragraph, the AdS/CFT curves  are considerably broader than the
corresponding kinetic theory results  for $\RR > 5$.

Examining \Fig{EPcompare}   a striking feature of the AdS result 
is  the dramatic transition from hydrodynamic behavior at $\RR=5$
to vacuum physics at $\RR=1$ which is not present in the kinetic theory
calculation.  This transition was noted previously and suggested as 
a way to reveal the strong coupling dynamics experimentally \cite{Noronha:2008un}.  
However, the absence of this transition in the weak coupling calculation
reflects a limitation of the kinetic theory approximation to QCD rather than 
a distinguishable difference between the AdS/CFT correspondence and weakly
coupled QCD.  Certainly the quantum  dynamics at $R \sim 1/\pi T$ can not be
captured by the semi-classical kinetic theory results.   It would 
be interesting to compute the stress tensor in this region in 
fixed order finite temperature perturbation theory to see if the dynamics of the two theories is similar at these length scales. As 
discussed in \Sect{comparison} using a short distance expansion \cite{Gubser:2007nd,Yarom:2007ap}, in the AdS/CFT calculation 
new length scales emerge at distances of order $1/\gamma^{1/2} \pi T$ and $1/\gamma^{3/2} \pi T$ which are not visible  in \Fig{EPcompare}. These scales have
been associated with saturation physics \cite{Hatta:2008tx,Dominguez:2008vd}.

Finally, we have analyzed the transition to the hydrodynamic regime in kinetic
theory and the AdS/CFT. In particular we have determined the source appropriate
for first and second order hydrodynamics in  each theory, following in part a
method outlined by Chesler and Yaffe \cite{Chesler:2007sv}.  This is elaborated
in \Sect{hydro:sec}, and the precise source (which  takes the form of 
derivatives of delta functions acting at the position of
the quark) is given in Table~\ref{sourcecoeff}.  
The source is constructed so that
the hydrodynamics to a given order, together with a source  at the same 
order, reproduces the  stress tensor of the full theory up to higher  order
powers of $\ell_{\rm mfp}/R$. \Fig{fitfigure} given in \app{details} 
fits our extracted hydrodynamic source with a polynomial at small $\k$ 
and the superb agreement with our full numerical results justifies the source analysis of \Sect{hydro:sec}.

\Fig{Boltzh} compares first order and second order hydrodynamics to the kinetic
theory results. Generally the second order theory provides only a minor
improvement to the first order results  until rather large radii, $\RR \gsim
40$.  Indeed the behavior of the second order theory seems rather unphysical
for $\RR \lsim 10$.  This shows the limitations of second order hydrodynamics.
Second order hydrodynamics
is constructed to reproduce the full results order by order at 
asymptotic distances and is not constructed to describe the decay 
of non-equilibrium transients produced by the  heavy quark.

The slow convergence of hydrodynamic expansion to the full results  of kinetic
theory  can be contrasted with the rapid convergence seen in the  AdS/CFT
results in \Fig{AdSh}.  In the AdS/CFT case  we see that  first (second)
order hydrodynamics describes the full result  at the  20\% (4\%) level for
$\RR \simeq 5$.  The agreement with hydrodynamics is not as good as described
earlier by Chesler and Yaffe.  This is because we are describing a quark moving
with velocity $v=1$, and we have found that the deviation from equilibrium is
noticeably larger than the $v=0.75$ quarks studied by these authors. In addition
the energy density distributions show larger deviations from first order
hydrodynamics and were not studied previously.  However, once (important) second order hydrodynamic
corrections  are included, the agreement with hydrodynamics is
remarkable already at modest $\RR$.

 We should mention that we have used the ``static" version
of second hydrodynamics  which  specifies $\pi^{\mu\nu}$ with 
a constituent  relation analogous to the first order constituent relation \cite{Baier:2007ix}.
Israel-Stewart type equations rewrite and interpret the constituent relation
as a dynamic equation  using lower order equations  of motion.
This  renders the system of 
equations hyperbolic and causal, but mixes orders in the gradient expansion. 
We have compared the static and the dynamic theories  for the kinetic and AdS
theories in \Fig{2ndhydro1}. Generally the Israel-Stewart type 
resummations do not lead to a significant improvement. Indeed, at smaller $\RR$
than shown in \Fig{2ndhydro1}\, Israel-Stewart type resummations
can lead to spurious shocks which are not reproduced by the full result. 
The differences between the static and dynamic theories gives
an estimate of higher order terms,  and this difference is smaller 
in AdS/CFT than in kinetic theory at the same $\RR$.

Clearly, the convergence to the hydrodynamic limit 
is significantly faster in the $\N=4$ theory relative to 
kinetic theory, even when lengths are measured in the scaled units 
described by $\RR$.  We remark that in the AdS/CFT 
the second order hydrodynamic parameter $\tau_\pi$ 
is a factor of $2.5$ smaller in scaled units than the   corresponding  kinetic
theory parameter,
\begin{align}
\frac{\tau_\pi}{\eta/sT} =& 6.32 \, ,      & \mbox{(Kinetic Theory)} \\
 \frac{\tau_\pi}{\eta/sT } =&   4  - 2 \log(2) \simeq 2.61 \, . &\mbox{(AdS/CFT)}
\end{align}
Based on these coefficients, it is natural to expect that the convergence
to the hydrodynamic limit is faster for the $\N=4$ theory  than the 
corresponding kinetic theory.
In theories based on quasi-particles and kinetic theory  it is difficult
to reduce the value of $\tau_{\pi}$ in scaled units significantly \cite{York:2008rr}.
Thus,  it would seem that our principal result of this 
study is reasonably generic. Specifically, 
based on the model theories  studied in this work we expect theories
without quasiparticles to approach  the hydrodynamic limit  several times
faster (in scaled units) than theories  based on a quasiparticle description.
From a practical perspective of applying hydrodynamics to various 
almost equilibrium phenomena of heavy ion physics (\emph{e.g.} the hydrodynamic flow due to jets and other local disturbances) this factor of two can be quite important. \\
{}\\
\noindent{\bf Acknowledgments:} \\
{}\\
P.~Chesler is supported by a Pappalardo Fellowship at MIT.
D.~Teaney and J.~Hong are supported in part by the Sloan Foundation and 
by the Department of Energy through the Outstanding Junior Investigator 
program, DE-FG-02-08ER4154.

\appendix

\section{The kinetic theory source in a leading-log approximation}
\label{source_app}

 The source of non-equilibrium gluons arises as gluons scatter off the 
heavy quark, $g  + {\rm Q} \rightarrow g + {\rm Q}$. The squared matrix element  for 
this process is  
\st
 \left|\M \right|^2 = \left[\frac{g^4 C_F N_c}{2d_A} \right] 16 \left[ \frac{2 (K\cdot P)^2}{Q^4} -\frac{ M^2 }{Q^2} +  \frac{M^2}{4 (K\cdot P)^2 } \right] \,  ,
\stp
where $K$ is the heavy quark momentum, $P$ is the gluon momentum, $Q = P'-P$ 
is the four momentum transferred to the gluon,  and we have averaged over 
the colors and spins of the external gluon.
In a leading log approximation 
only the (first) most singular term  is kept.
The source of non-equilibrium gluons of momentum $\p$ is obtained from the Boltzmann collision integral for the  $g +Q \rightarrow g + Q$ process
\begin{align}
S(t,\x,\p)=&S(\p)\delta^3(\x-\v t)\nonumber\\
=&
-\int_{kp'k'}\frac{|\M|^2}{16 k^0 k^{'0} p p'} (2\pi)^4\delta^4(P_{tot})
[f_pf_k(1+f_{p'})(1+f_{k'})-f_{p'}f_{k'}(1+f_p)(1+f_k)] \, ,
\end{align}
where the heavy quark distribution $f_k=(2\pi)^3\delta^3(\k-\k_H)\delta^3(\x-\v t)$ is out of equilibrium.  

We expand the source  in a spherical harmonic basis 
in the $(x,y,z)$ coordinate system
\begin{align}
S(\p)=&\sum_{l,m}S_{lm}(p)H_{lm}(\hat{\p};zx) 
 =\sqrt{\frac{2l + 1}{4\pi}} \sum_l S_{l0}(p) P_l(\cos\theta_{\p\k}) \, ,
\end{align}
and note that the $S_{lm}$ vanishes for non-zero $m$ due to the 
azimuthal symmetry of the problem.
Using the orthogonality of $P_{l}(\cos\theta_{\p\k})$ and  the  phase-space
parameterization and kinematic approximations used to analyze the energy loss of
heavy quarks \cite{Moore:2004tg},  the expansion coefficients can be
written \footnote{
Specifically, we use Eq.~(B21) of \Ref{Moore:2004tg}. However, we have interchanged
the role of $\p$ and $\k$ to be consistent with the 
notation used in this work.  
 }
\st
S_{l0}(p)= -\sqrt{\frac{2l + 1 }{4\pi}}\int_0^\infty dq \int_{-vq}^{vq} \frac{d\omega}{v} 
\int\frac{d\phi}{2\pi} P_{l}(\cos\theta_{\p\k}) 
\frac{|\mathcal{M}|^2 }{16 p^2 (k^0)^2 }
\left[f_p(1+f_{p+\omega})
-f_{p-\omega}(1+f_p)\right] \, ,
\stp
where $\omega$ is the energy transfer, $\q = \p' - \p$ is the three momentum
transfer, and $\phi$ is the azimuthal angle.  The matrix elements in this parameterization are
\st
\frac{|\mathcal{M}|^2 }{16 p^2 (k^0)^2 }
= \left[\frac{g^4  C_F N_c}{2d_A} \right] \frac{2 (1 - v\cos\theta_{\k\p} )^2 } {(q^2 - \omega^2)^2}  \, ,
\stp 
where $\cos\theta_{\k\p}$ is expressed in terms of the integration
variables, $\omega$, $q$ and $\phi$ \cite{Moore:2004tg}.
Now we  consistently expand out the integrand to quadratic order in 
$\omega/T$ and $q/T$. This includes three types of terms:  (1) an  
expansion of the distribution functions to quadratic order,  (2)
an expansion of the angle $\cos\theta_{pk}$ 
to linear order in $q/T$ 
  and   (3) an expansion of the Legendre polynomial to linear
order, $P_l(x + \delta x) \simeq P_{l}(x) + P_{l}'(x)\, \delta x$.
With the full expansion we explicitly integrate over the azimuthal 
angle $\phi$ and the energy $\omega$,   observing empirically that all harmonics vanish for  $l >1$  when the velocity
is lightlike.  
Further,  the $l=0$ and $l=1$ harmonics  can be done analytically 
leading to a simple answer recorded in \Eq{source} 
\begin{subequations}
\bg
S_{00}(p)
&=&\frac{1}{\sqrt{4\pi} T^2}\, \left[\frac{g^4C_F N_c}{2d_A} \right] \, \log\left(\frac{T}{m_D}\right)
f_p(1+f_p)\left(-\frac{2T }{p}+ 1+2f_p \right) \, , \\
S_{10}(p)
&=&\frac{1}{\sqrt{12\pi} T^2} \, \left[ \frac{g^4 C_F N_c}{2d_A} \right] \, \log\left(\frac{T}{m_D}\right)
f_p(1+f_p) (1+2f_p) \, ,
\nd
\end{subequations}
The leading-log simplifications
described in this paragraph were observed previously when computing
the shear viscosity.

\section{Numerical Details about Kinetic theory and the Fourier Transform}
\label{details}

The goal of this appendix is to give some of the details of how the stress
tensor is computed in kinetic theory.  Most of the notation and strategy
follows an appendix of \Ref{Hong:2010at} and this reference should be consulted
for the full details.  

The  linearized Boltzmann equation in Fourier space reads
\st
\label{linboltz}
(-i\omega+i\v_p\cdot\k)\delta f(\omega,\k,\p)
=C[\delta f,\p]+ 2\pi S(\p)\delta(\omega-\v\cdot\k) \, ,
\stp
where $\v_\p=\hat\p$ is the particle velocity and the vector $\k$ in the laboratory coordinate system is 
\st
 \k = (k^x, k^y, k^z) = k (\sin\theta_\k \cos\varphi_k, \, \sin\theta_\k  \sin\varphi_k, \,  \cos\theta_\k) \, .
\stp
In order to solve \Eq{linboltz} numerically,
it is convenient to introduce the Fourier coordinate system $(x',y',z')$ 
where $\hat{\z}'$ points along the Fourier momentum $\k$
\begin{align}
   \hat \x' =& \frac{k}{k_T}  \hat{\k} \times (\hat{\v} \times \hat \k) \, ,  \\
   \hat \y' =& \frac{k}{k_T}  \hat{\v} \times \hat \k \, , \\
   \hat \z' =& \hat \k  \, .
\end{align}
Following our previous work \cite{Hong:2010at} we re-express the source 
and ultimately the solution $\delta f$ in 
terms of real spherical harmonics with respect to the 
$\hat{\x}',\hat{\y}', \hat{\z}'$  coordinate system:  
\st
    H_{lm}(\hat \p;z'x') = N_{lm} P_{l|m|}(\cos\theta_\p) \times 
\left\{
\begin{array} {ll}
1    & \mbox{\quad for $m=0$} \\
\sqrt{2}  \cos m\varphi_\p &\mbox{\quad for $m > 0$ } \\
\sqrt{2}  \sin |m|\varphi_\p &\mbox{\quad for $m < 0$ } \\
\end{array} 
\right.  \, ,
\stp
where $N_{lm}$ is a normalization factor \cite{Hong:2010at}. 
We note that the unit vector  $\hat\p$ has the following
components
\st
\hat{\p}^{x'} = \sqrt{ \frac{4\pi}{3} } H_{11}(\hat\p;z'x')\, , \qquad \hat{\p}^{y'} = \sqrt{\frac{4\pi}{3}} H_{1,-1}(\hat \p;z'x') \, , \qquad
   \hat{\p}^{z'} = \sqrt{\frac{4\pi}{3}} H_{10}(\hat \p;z'x')\, . 
\stp

Since the distribution function $\delta f$ 
is independent of 
the azimuthal angle of $\k$   with respect to the  original 
$x,y,z$ coordinate system, we will  choose this azimuthal angle 
$\varphi_k$ to 
  be zero so that $\k$ lies in 
the $x,z$  plane. Then in the $(x',y',z')$ coordinate system the vector
$\v$ has the components
\[
\hat{\v} = (v^{x'}, v^{y'}, v^{z'} ) =  (\sin\theta, 0, \cos\theta) \, ,
\]
and  
\st
 \hat{\v} \cdot \hat{\p} =  \sqrt{ \frac{4\pi}{3} } \Big( 
 \cos\theta \, H_{10}(\hat \p ; z'x') + \sin\theta \,  H_{11}(\hat \p; z'x') \Big) \, .
\stp
%
%

The  steady state solution to the linearized Boltzmann  equation is also expanded in the spherical harmonics defined above
\st
  \delta f(\omega,\k,\p) =  \sum_{lm} 2\pi \delta(\omega - \v \cdot \k) n_p (1 + n_p) \chi_{lm}(p,\k) H_{lm} (\hat{\p};z'x') \, .
\stp
and the Boltzmann equation for $\chi_{lm}$ becomes 
\begin{align}
 \left(-i\omega \delta_{ll'}  +
ik C_{ll'}^{m}  \right) p^2 n_p(1+n_p) \chi_{l'm}
=   C_{lm}[\delta f,\p]  +  {\mathcal H} p^2 n_p(1+ n_p) \frac{\mu_A}{T} {\mathcal S}_{lm}(p, \theta)  \, , 
\end{align}
where the $m$ index is not summed over.
Here $C_{ll'}^m$ is a Clebsch Gordan coefficient \cite{Hong:2010at}, $C_{lm}[\delta f,\p]$ 
is the collision integral in this basis, 
the normalization coefficient  is 
\st
  {\mathcal H} = \frac{\mu_{F}}{T^3d_A\mu_A}  \, ,
\stp
and  the source is
\begin{multline}
\label{numerical_source}
{\mathcal S}_{lm}(p,\theta) = \frac{1}{2 \xi_B/T^3 }  \Big[ \left( \frac{-2T}{p}  + 1 + 2n_p \right) \sqrt{4\pi} \delta_{l0}\delta_{m0}   \\
    +\sqrt{\frac{4\pi}{3} } \left( 1 + 2 n  \right) \left(\delta_{l1}\delta_{m0}\cos\theta + \delta_{l1} \delta_{m1} \sin\theta \right) \Big] \, .
\end{multline}
For each value of $(k_x,k_z)$ (in units of $\mu_{A}/T$)  
the linear equations are solved  for $F_{lm} \equiv \chi_{lm}/{\mathcal H}$. 
Due to rotational 
invariance of the collision operator around 
the $\k$ axis,  the  matrix equation does not mix
harmonics  with different magnetic quantum numbers, {\it i.e.} the collision
operator is diagonal in  $m$. 
Thus, the matrix equation in $l,l'$ is solved for $m=1$ and $m=0$ separately. 
Harmonics with $m > 1$ are not sourced by the motion of the quark
in a leading log approximation.

After solving for $\delta f(\omega,\k,\p)$, the energy and momentum excess due 
to the moving quark can be computed using the kinetic theory
\st
\delta T^{0\mu}(\omega,\k)=2d_A\int \frac{\dd^3\p}{(2\pi)^3} \,  p^\mu \delta f(\omega,\k,\p) \, , 
\stp
where $T^{0\mu}(\omega,\k)$ is proportional to $2\pi \delta (\omega - \v\cdot \k)$
\st
   \delta T^{0\mu}(\omega, \k) \equiv 2\pi \delta(\omega-\v\cdot \k) \tilde T^{0\mu}(k_z,k_T)  \, .
\stp
The relationship between the $(x,y,z)$ and the $(x',y',z')$ coordinate system is
\begin{align}
\delta T^{0x}(\omega,\k) =& \cos\theta \, \delta T^{0x'}(\omega,\k) - \sin\theta \, \delta T^{0z'}(\omega,\k)  \\
\delta T^{0y}(\omega,\k) =& 0  \\
\delta T^{0z}(\omega,\k) =& 
\sin\theta \, \delta T^{0x'}(\omega,\mathbf{k}) +\cos\theta \, \delta T^{0z'}(\omega,\mathbf{k}) \, ,
\end{align}
In presenting these formulas  we have taken $\k$ in the $x,z$ plane,
\emph{i.e.} $\varphi_k = 0$.
More generally rotational invariance dictates that  $(T^{0x}, T^{0y})$ 
is proportional to $\hat \k_T$
\begin{align}
  \delta \tilde T^{0x}(\k)  =& \delta \tilde T^{0k_T}(k_z,k_T) \cos\varphi_k  \, ,  \\
  \delta \tilde T^{0y}(\k)  =& \delta \tilde T^{0k_T}(k_z,k_T) \sin\varphi_k  \, ,
\end{align}
and the preceding discussion with $\varphi_k=0$ suffices to determine $T^{0k_T}(k_z,k_T)$.

The stress tensor is tabulated in the $k_z, k_T$ plane 
and then Fourier transforms can be used to compute the stress
tensor in coordinate space 
\st
 \delta T^{0\mu}(t,\x) = \int_{-\infty}^{\infty} \frac{\dd\omega}{2\pi} \int \frac{\dd^3\k}{(2\pi)^3} e^{-i\omega t + i \k \cdot \x } \, \delta T^{0\mu}(\omega, \k) \, .
\stp
Employing the familiar identity
\st
e^{i k_T x_T \cos(\varphi_r - \varphi_k)} = J_{0}(k_T x_T) + 2\sum_{n} i^{n} J_{n} (k_T x_T) \cos(n(\varphi_r - \varphi_k) ) \, , 
\stp
it is not difficult to show that 
\begin{subequations}
\label{fourierints}
\begin{align}
 \delta T^{00}(x_L, x_T) 
=&   \int_{0}^{\infty} 
\frac{k_{T} \dd k_{T}}{2\pi}  J_{0}(k_{T} x_T) 
\,
\int_{-\infty}^\infty \frac{\dd k_{z}}{2\pi} e^{ik_z z} \, \delta \tilde T^{00}(k_z,  k_T) \, , \\
\label{fourierints2}
\delta T^{0x_T}(x_L, x_T) =&  
  \int_0^{\infty} \frac{k_T \dd k_T}{2\pi}  J_{1}(k_T x_T)  \int_{-\infty}^{\infty} \frac{\dd k_z}{2\pi} e^{ik_z z}  i \, \delta \tilde T^{0k_T} (k_z,k_T)  \, , \\
\delta T^{0z}(x_L, x_T) =&  
  \int_0^{\infty} \frac{k_T \dd k_T}{2\pi}  J_{0}(k_T x_T)  \int_{-\infty}^{\infty} \frac{\dd k_z}{2\pi} e^{ik_z z}  \, \delta \tilde T^{0k} (k_z,k_T)  \, .
\end{align}
\end{subequations}


The Fourier integrals in
Eq.~(\ref{fourierints}) are not particularly easy.  In
order to get  a convergent integral, we first multiply the numerical data by
a window function which eliminates the contributions of high frequency modes. 
For kinetic theory results,   a  sample window  function is 
\st
W(k)= \frac{1}{2}\bigg[1-\mbox{erf}\left((k -k_{\rm max})/\sigma\right)\bigg].
\stp
with $k_{\rm max} = 7.5 \mu_{A}/T$ and $\sigma = 3.5 \mu_{A}/T$
while for the AdS/CFT the $k_{\rm max}$ and $\sigma$ was considerably larger, $k_{\rm max} = 80 \pi T$ and $\sigma =60\pi T$.
We computed the integrals in Eq.~(\ref{fourierints}) 
using two methods.
The first method used  brute force  summation to compute these integrals.
In the second method  we 
re-parameterized the  stress tensor by the source functions
$\phi_v$ and $\phi_k$  used in the hydrodynamic analysis.
Specifically, we \emph{define} the functions $\phi_v(\k)$ and
$\phi_k$ by the equations
\begin{align}
\label{phivphikdef}
\delta  T^{0x'}(\omega,\k)
=&\frac{i\sin\theta\phi_v(\k)}
{\omega+iD k^2} \, 2\pi \mu_F \, \delta(\omega-k\cos\theta) \, , \\
\delta T^{0z'}(\omega, \k)
=&\frac{i \left(\omega\cos\theta\phi_v(\k)+c^2(k)k\right)-k\omega\phi_k(\k)}
{\omega^2-c^2(k) k^2+i\Gamma_s\omega k^2} \, 
2\pi\mu_F \, \delta(\omega-k\cos\theta) \, ,
\end{align}
where $c^2(k)=c_s^2(1+\tau_\pi\Gamma_sk^2)$ and $T^{00}(\omega,\k)$ is 
determined by energy-momentum conservation. This reparametrization is used
for all values of the Fourier momentum $\k$. Such a reparametrization is useful 
because, in contrast to $T^{0x'}$ and $T^{0z'}$, 
the functions $\phi_v(\k)$ and $\phi_k(\k)$ are smooth and can be easily and accurately interpolated. 
Then the integrals are computed using Gaussian quadrature 
with break points at the hydrodynamic poles.
The brute force summation and the more sophisticated numerical scheme give  the same answer at the 1\% level and are independent of the cutoff parameters.

\section{Details on the Hydrodynamic Source in Kinetic Theory and the AdS/CFT} 
\label{source_details}

The purpose of this appendix is to explain in somewhat greater detail
how the coefficients given in Table~\ref{sourcecoeff} are computed  for
both kinetic theory and the AdS/CFT. 
In the process, we will exhibit several fits to our numerical results. 
The quality of these fits indicates that the deviation of the stress 
tensor from its hydrodynamic form at small $\k$ and $\omega$  
is well described by a multivariate polynomial and justifies
the hydrodynamic analysis of \Sect{hydro:sec}.  We will first discuss
the AdS/CFT theory  and then indicate how the analysis can be applied to kinetic theory.

\subsection{AdS/CFT} 

The applicability of second order hydrodynamics to the AdS/CFT has been
questioned based on the analytic structure of retarded stress tensor
correlators in the lower half plane \cite{Noronha:2011fi,Denicol:2010xn}.  However, provided second order
hydrodynamics is used to describe the behavior of hydrodynamic modes ({\it i.e} a pole
in the retarded Green function arbitrarily close to the real axis),  rather than to model the decay of
non-hydrodynamic modes or transients 
({\it i.e.} additional analytic structure  in the lower half plane), second
order hydrodynamics is applicable to this strongly coupled theory.
This appendix serves to clarify this  points. 

After computing the exact stress tensor of the full AdS/CFT theory we {\it
define} the functions  $\phi_v(\k)$ and $\phi_k(\k)$ as in \Eq{phivphikdef}.
This would seem to be simply a reparametrization of the original numerical
data on $T^{0x'}$ and $T^{0z'}$ with two functions $\phi_v(\k)$ and
$\phi_k(\k)$.  However, as we will see, the functions $\phi_v(\k)$ and
$\phi_k(\k)$ are analytic functions of $\k$  while the original fields have 
poles arbitrarily close to the real axis as $\k \rightarrow 0$, see
\Eq{hydrosol}.  The purpose of first and second order hydrodynamics is to
describe the location of these poles.  First order hydrodynamics determines 
the pole location
to first order in  $k \ell_{\rm mfp}$,
but neglects terms of order $(k \ell_{\rm mfp})^2$ which are captured by second
order hydrodynamics\footnote{$\ell_{\rm mfp}$ should be taken as $1/\pi T$ in
the strongly coupled theory.}.  It should be emphasized that the pole shift is
a consequence of  modifying the ideal hydrodynamic {\it equations of motion} by
powers of $k \ell_{\rm mfp}$ rather than modifying the source.  Since, the
ideal solution has a hydrodynamic pole at $\omega=c_s k$ modifying the
equations of motion does not simply correct the {\it solution} by simple
powers $k\ell_{\rm mfp}$ close to the pole.
We  will determine the source functions $\phi_v(\k)$ and $\phi_k(\k)$ using
first and second order hydrodynamics. Specifically, we determine $\phi_v(\k)$
and $\phi_k(\k)$ using \Eq{phivphikdef} (with the same numerical data for the full stress tensor $T^{0x'}$ and $T^{0z'}$),  but  in the
first order case we set the second order transport coefficient to
zero in these equations\footnote{We note that $\phi_v$ 
is the same for first and second order hydrodynamics. Only $\phi_k$ is
affected by non-zero $\tau_\pi$.}, $\tau_\pi \rightarrow 0$. 

\begin{figure}
\begin{center}
\includegraphics[width=0.49\textwidth]{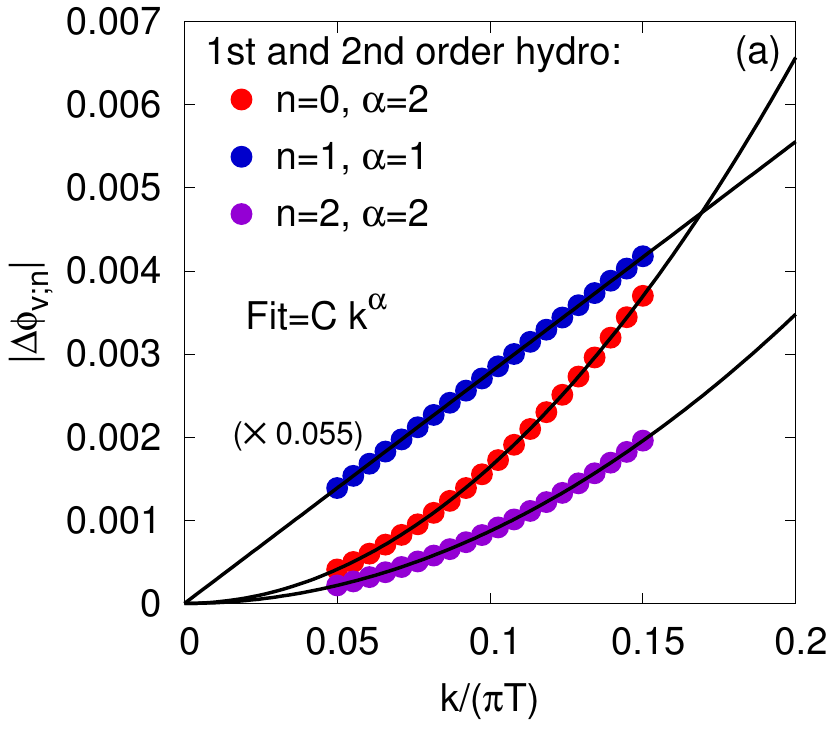}
\vskip 0.1in
\includegraphics[width=0.49\textwidth]{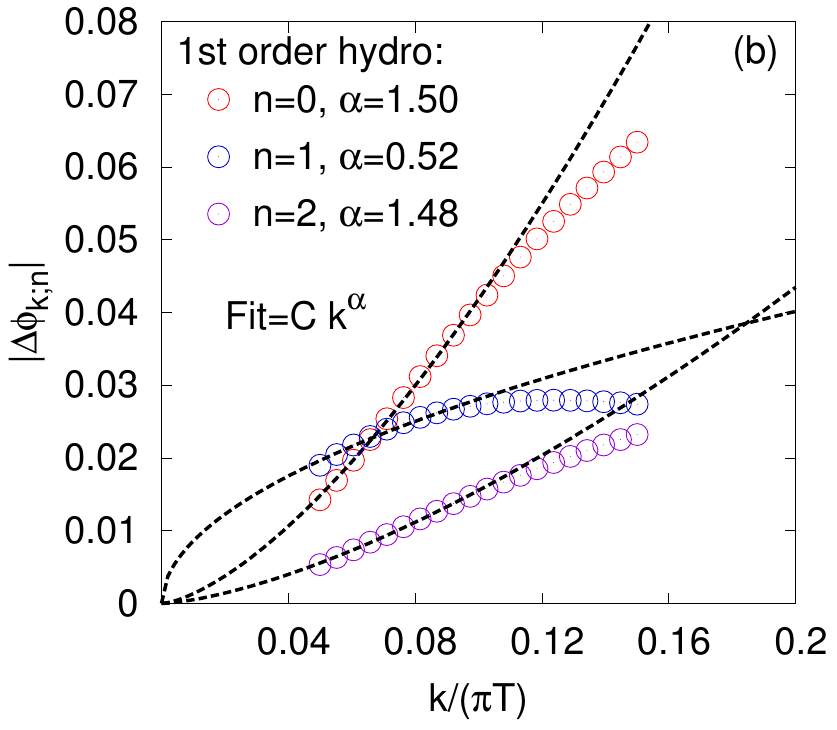}
\includegraphics[width=0.49\textwidth]{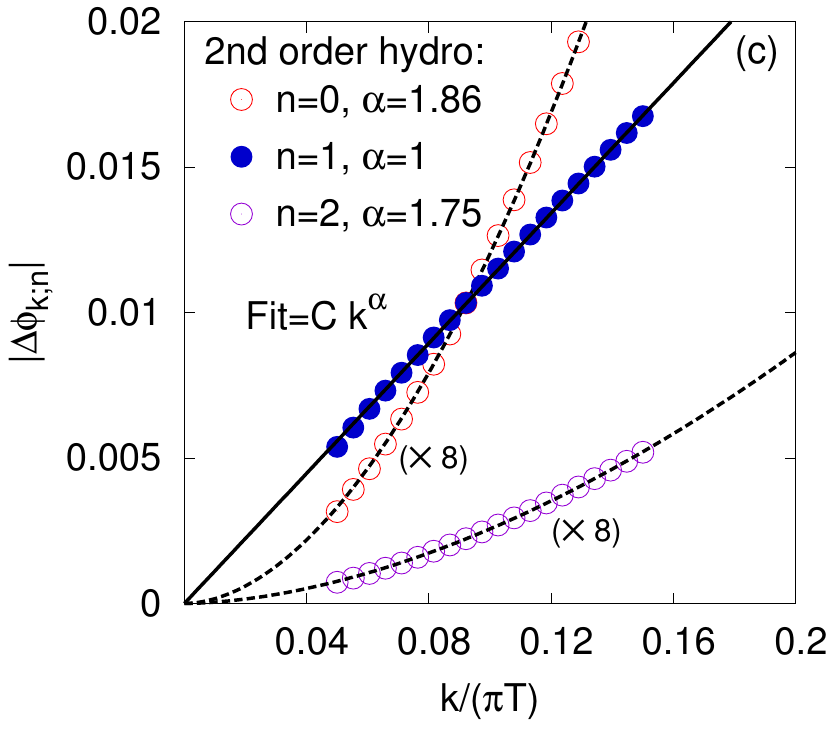}
\end{center}
\caption{
\label{adssourcefig}  
Hydrodynamic fits to the AdS/CFT source functions  $\phi_v \equiv 1 + \Delta \phi_v$ and $ \pi T \phi_k  = -\frac{1}{3}  +  \Delta \phi_k$ 
at small $k$ (see the text surrounding \Eq{phiv2defs}). 
The Fourier 
coefficients displayed in this  figure have been multiplied
by a factor indicated in parentheses to  increase visibility,
and  are fit with the functional form $C k^{\alpha}$.
The dashed lines and open symbols have non-integer fit-powers 
and lie beyond the description of hydrodynamics to the 
specified order, {\it i.e.} the fit is not expected to work.
(a) The $n=0$, $n=1$, and $n=2$ Fourier coefficients of $\phi_v$.
(b) The $n=0$, $n=1$, and $n=2$ Fourier coefficients of $\phi_k$,
when $\phi_k$ is extracted using first order hydrodynamics.
(c) Same as (b), but for second order hydrodynamics.
}
\end{figure}

Since the  source functions $\phi_v$ and $\phi_k$ are functions
of  $k$ and $\omega = k \cos\theta$, we can expand these functions in 
Fourier series
\begin{align}
\label{phiv2defs}
 \phi_v(\kn,\cos\theta) \equiv \phi_{v;0}(\kn) + 2\phi_{v;1}(\kn) \cos\theta + 2 \phi_{v;2}(\kn) \cos2\theta + \ldots .\\
 \pi T \phi_k(\kn,\cos\theta) \equiv \phi_{k;0}(\kn) + 2  \phi_{k;1}(\kn) \cos\theta + 2  \phi_{k;2}(\kn) \cos2\theta + \ldots .
\end{align}
In \Fig{adssourcefig}  we plot the terms of the Fourier series  $\phi_{v;n}(k)$ and $\phi_{k;n}(k)$ and fit these functions with a simple power law, $k^{\alpha}$.  

As we see from the fit,  $\phi_v$ is well 
described by  a quadratic polynomial  
at small $k$
\st
\label{phivfit}
   \phi_{v;0}(k) =  1 - 0.1643 \left(\frac{k}{\pi T}\right)^2 \, , 
  \qquad \phi_{v;1} = 0.5 i \left(\frac{k}{\pi T}  \right) \, ,
  \qquad \phi_{v;2} = 0.0870 \left(\frac{k}{\pi T} \right)^2 \, .
\stp
For first order hydrodynamics 
the terms of quadratic order can be neglected.
Our numerical result for $\phi_{v;1}$ is nicely consistent with the first order analytic result \Ref{Chesler:2007an}.)

Now we examine $k \phi_k$ using first and second order hydrodynamics\footnote{
We discuss $k \phi_k$ instead of simply $\phi_k$ since the source for
hydrodynamics is $\phi_k(k) \k$.}. When first order hydrodynamics is used
the source function is not well described by a polynomial (see
\Fig{adssourcefig}(b)). However,  we see that $k \phi_k$  decreases faster than $k$ (as $k^{1.52}$ for
$n=1$),  and therefore $k\phi_k$ can be neglected in a first order hydrodynamic
analysis of the long wavelength response to the heavy quark.  A 
concerned reader might worry that the resulting source function $k \phi_k$ is not well described by a quadratic polynomial, and incorrectly conclude  a local
source for hydrodynamics can not be constructed beyond linear order.
However, when second order hydrodynamics is used to extract the source through quadratic order (as is
appropriate), $k \phi_k$ {\it is} well described by the
quadratic polynomial  -- see the linear fit in \Fig{adssourcefig}(c) for $\phi_{k;1}$.
Numerically we find
\st
\label{eqphikfit}
  \phi_{k;0} = -\frac{1}{3}\, ,  \qquad   \phi_{k;1} = 0.11i \frac{k}{\pi T} \, ,
\stp
up to non-analytic terms that fall faster than $k^2$.   (These 
non-analytic terms could be removed by pushing
the hydrodynamic analysis to third order.)
In summary, we see that using second order hydrodynamics neatly removes the non-analytic behavior seen in the first order source, and then the  source 
for second order  hydrodynamics is then well described by a quadratic polynomial. 
The  coefficient  $\tau_\pi$ which shifts the hydrodynamic pole is universal --  it was determined through an analysis of  two point functions 
\cite{Baier:2007ix},
and the same coefficient determines the long distance response to the 
disturbance produced by a heavy quark.  By contrast 
the source functions $\phi_v$ and $\phi_k$  are not universal but depend on the particular
way in which the heavy quark couples to hydrodynamic modes. Using the 
fits in \Eq{phivfit} and \Eq{eqphikfit} (and the relation between
$\phi_v,\phi_k$ and $\phi_1$,$\phi_2$ given in \Eq{Shydroexpand}), we parametrize this
source by three numbers to  quadratic order which are given in Table~\ref{sourcecoeff}.

\subsection{Kinetic Theory}

\begin{figure}
\begin{center}
\includegraphics[width=0.495\textwidth]{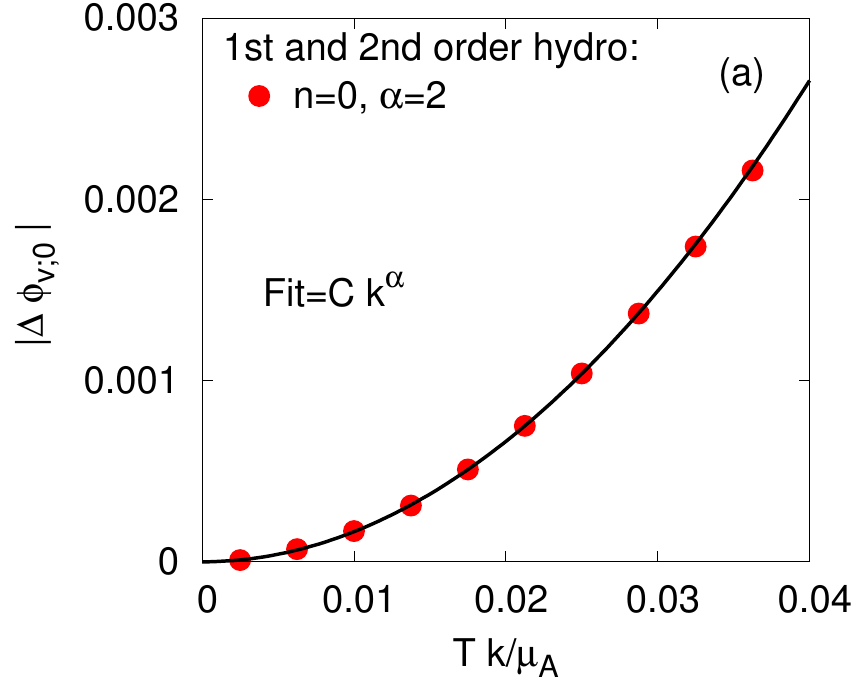} 
\includegraphics[width=0.495\textwidth]{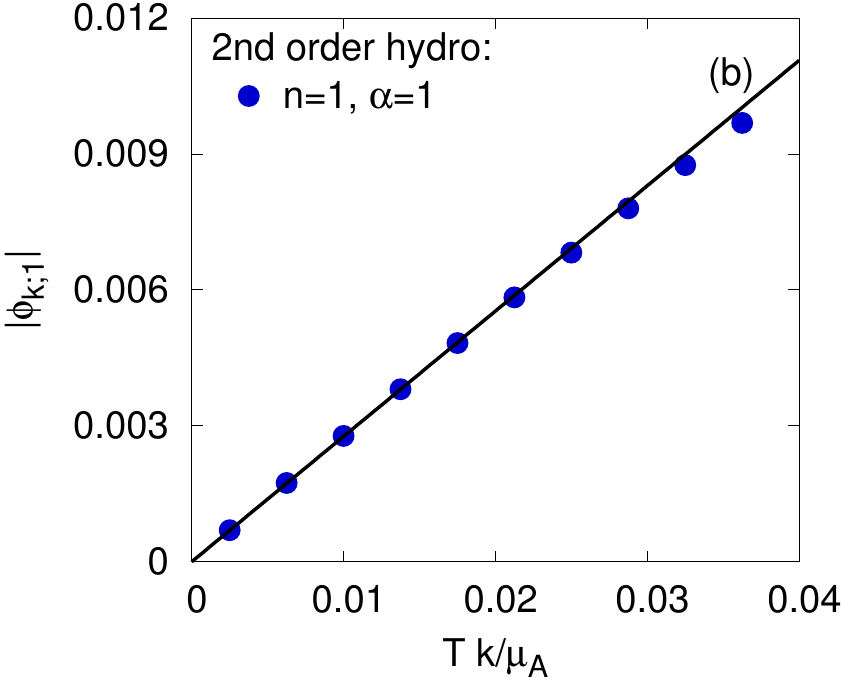} 
\end{center}
\caption{ 
\label{fitfigure}
(a) A polynomial fit to the source function $\phi_{v;0}$ defined in \Eq{phiv2defs}. The fit result is recorded in \Eq{fitresult}.  (b) Using the fit coefficient from $(a)$, a prediction
is made $\phi_{k;1}$ (see \Eq{fitresult}) using the analysis of  \Sect{hydro:sec}.
\label{hydrofitplot}
}
\end{figure}

As in the previous appendix, we {\it define} the functions
$\phi_v$ and $\phi_k$ from the exact stress tensor (\Eq{phivphikdef}),  and again 
expand these functions in 
Fourier series as in \Eq{phiv2defs}, but, as is appropriate for the kinetic theory calculation, $\pi T \phi_k$ is replaced with
$(\mu_A/T) \phi_k$ .
Then the Fourier coefficients  are fit  with a polynomial which are shown in
\Fig{fitfigure}.  It was verified that  the other Fourier coefficients of $\phi_v$ and $\kn \phi_k$  that are not  shown decrease faster than $\kn^2$ and thus lie outside of the hydrodynamic analysis  which is restricted to second order
inclusive. 
Examining the fits shown in \Fig{hydrofitplot}, we see that  the parameterization
\begin{align}
\label{fitresult}
    \phi_{v;0} =& 1 +  1.66(1) \left( \frac{k T}{\mu_A } \right)^2 \, , \\
    \phi_{k;1} =& -\frac{1}{6} 1.66(1) i  \frac{k T}{\mu_A} \, ,
\end{align}
describes our  numerical data at small $k$ and $\omega=k\cos\theta$  
quite well.  The fact that $\phi_{v;0}$ and $\phi_{k;1}$ have the same fit 
coefficient (up to a symmetry factor of $1/6$) 
is a consequence of the hydrodynamic analysis  of \Sect{hydro:sec}.
Comparing the fit with the functional form given in the text (\Eq{Shydroexpand}), 
 we see that we have a non-zero $\phi_{2}^{(0,0)}$ (which is recorded in Table ~\ref{sourcecoeff}),
but the coefficients $\phi_{1}^{(0,0)}$ and $\phi_1^{(1,0)}$
seem to vanish.  In fact $\phi_1(\omega,k^2)$ vanishes   to all orders in $\omega,k$ as we will now show.

To this end we can return to the 
analysis given in \Sect{hydro:sec}. 
For a given $\k$, we expect that there is a non-zero  component of $T^{ij}(\omega,\k)$
which transforms 
as a spin two tensor under  azimuthal rotations around the $\k$ axis.
Examining the decomposition of $\tau^{ij} \equiv T^{ij} - T^{ij}_{\rm hydro}$ 
into tensor structures (\Eq{tauijbasisfcns}),
and noting that hydrodynamics does not yield
such a  spin two tensor, 
we see that this spin two 
component of $T^{ij}$ determines 
$\left[v^iv^j-\frac{1}{3}v^2\delta^{ij}\right] \phi_1$, since this is
the only tensor structure of $\tau^{ij}$  
that has a spin two component under azimuthal rotations around $\k$.
Examining the source for the Boltzmann equation given in 
 \Eq{numerical_source} of \app{details} we see that the specific form of
the  source does not excite the spin two (i.e. $m=2$) 
components of the distribution function $\delta f$.  
Thus, since the Boltzmann equation does not mix harmonics of different spin, 
the spin two component of the stress tensor 
vanishes and so does $\phi_1$.  This approximate symmetry 
is specific to the simplified form of the source which arises in 
a leading-log approximation and is not expected to hold more generally.
 

\bibliography{HQbib}


\end{document}